\UseRawInputEncoding
\documentclass[twocolumn]{aastex631}

\usepackage{amsmath}
\usepackage{graphicx}
\usepackage[caption=false]{subfig}
\usepackage{appendix}
\usepackage{float}

\newcommand{\bs}{\boldsymbol}

\shorttitle{GP Methods for Large Astrometric Data Sets}
\shortauthors{Hapitas et al.}

\begin{document}

\title{Gaussian Process Methods for Very Large Astrometric Data Sets}

\author[0009-0007-4319-2216]{Timothy Hapitas}
\affiliation{Department of Physics, Engineering Physics and Astronomy, \\
Queen's University, Kingston, Ontario, Canada \\
}

\author[0000-0001-6211-8635]{Lawrence M. Widrow}
\affiliation{Department of Physics, Engineering Physics and Astronomy, \\
Queen's University, Kingston, Ontario, Canada \\
}

\author[0000-0002-9583-5216]{Thavisha E. Dharmawardena}
\affiliation{ Center for Computational Astrophysics, \\
Flatiron Institute, New York, NY, USA \\
}

\author[0000-0002-9328-5652]{Daniel Foreman-Mackey}
\affiliation{
Center for Computational Astrophysics,
Flatiron Institute, New York, NY, USA
}

%\collaboration{20}{(AAS Journals Data Editors)}

\begin{abstract}
We present a novel non-parametric method for inferring smooth models of the mean velocity field and velocity dispersion tensor of the Milky Way from astrometric data. Our approach is based on Stochastic Variational Gaussian Process Regression (SVGPR) and provides an attractive alternative to binning procedures. SVGPR is an approximation to standard GPR, the latter of which suffers severe computational scaling with $N$ and assumes independently distributed Gaussian Noise. In the Galaxy however, velocity measurements exhibit scatter from both observational uncertainty and the intrinsic velocity dispersion of the distribution function. We exploit the factorization property of the objective function in SVGPR to simultaneously model both the mean velocity field and velocity dispersion tensor as separate Gaussian Processes. This achieves a computational complexity of $O(M^3)$ versus GPR's $O(N^3)$, where $M << N$ is a subset of points chosen in a principled way to summarize the data. Applied to a sample of $\sim 8 \times 10^5$ stars from the Gaia DR3 Radial Velocity Survey, we construct differentiable profiles of the mean velocity and velocity dispersion as functions of height above the Galactic midplane. We find asymmetric features in all three diagonal components of the velocity dispersion tensor, providing evidence that the vertical dynamics of the Milky Way are in a state of disequilibrium. Furthermore, our dispersion profiles exhibit correlated structures at several locations in $|z|$, which we interpret as signatures of the Gaia phase spiral. These results demonstrate that our method provides a promising direction for data-driven analyses of Galactic dynamics. 
\end{abstract}

\keywords{Galaxy: kinematics and dynamics --- methods: statistical --- astrometry}

%%%%%%%%%%%%%%%%%%%%%%%%%%%%%%%%%%%%%%%%%%%%%%%%%%%%%%%%%%%%%%%%%%%%%%%%%%%%%%

\section{Introduction} \label{sec:intro}

To an excellent approximation, the stellar disk of the Milky Way (MW) is a collisionless system whose dynamical state is described by a single-particle phase space distribution function (DF), $f({\bf x},{\bf v},t)$, which obeys the collisionless Boltzmann equation (CBE). Though $f$ fully specifies the state of the system, it is often more practical to work with velocity moments of $f$, which satisfy a sequence of partial differential equations in the spatial coordinates and time. For example, the evolution of the zeroth moment, that is, the number density, is given by the continuity equation. The time-derivative of the first moment is given by the Jeans equations, and depends on second moments of the DF and the gravitational potential. Velocity moments allow one to connect the theoretical framework of Galactic dynamics with observational data and are ubiquitous in the study of the Galaxy \citep{Binney2011, Eilers2019, Bovy2013, Sarkar2019, Wang2022}. \newline

The most common method to estimate velocity moments is to compute weighted averages of $v_i$, $v_i v_j$, etc within 3D bins or voxels from large astrometric survey data. This approach was applied to the Gaia Radial Velocity Survey (RVS) \citep{GaiaMission} by numerous authors to study the kinematics of the Milky Way (see, for example, \citep{Gaia2023, Gaia2018, Wang2022}). However, binning has drawbacks that weaken the connection between observation and theory. In particular, results depend on the bin size, which governs the balance between spatial resolution and the signal-to-noise ratio. Moreover, since the Jeans equations involve derivatives of the velocity moments, finite differencing of binned quantities is required to estimate the gravitational potential. An alternative approach is to use parametric models for the DF and the potential. For two early examples of the use of parametric models to estimate the vertical potential, see \citet{bahcall1984} and \citet{kuijken1989}.\newline

In this work, we present a non-parametric method for simultaneously inferring the mean velocity and velocity dispersion from astrometric survey data. Our starting point is the assumption that measurements of stellar velocities can be regarded as statistical draws from an underlying latent function, namely, the mean velocity field $\overline{\bs{v}}$. The scatter in these measurements arises from two effects: observational uncertainties and the intrinsic velocity dispersion of the DF. Since the latter, which is a function of position, plays a central role in the Jeans equations, we devise a method where it is also treated as a latent function. Our method is based on Gaussian Process Regression (GPR), a machine learning framework that is well-suited to the task of inferring an underlying latent function from a set of noisy observations. With GPR, one makes the additional assumption that the values of the latent velocity field in the MW form a Gaussian Process (GP). That is, one assumes any finite set of measurements and/or latent function values follows a multivariate normal distribution whose covariance function or kernel depends on the separation vector between measurement pairs. Inference in GPR is performed by optimizing the negative log likelihood of the regression model through gradient descent, providing values for the GP hyperparameters that best describe the data. One then constructs a posterior distribution over the latent function using these optimal hyperparameters, which serves as the prediction for the latent function. Previously, GPR was used by \citet{Nelson2022} to derive the velocity field in the vicinity of the Sun from the GDR2 RVS catalog. As an application, they mapped the divergence of $\overline{\bs{v}}$, which is related to the total time-derivative of the density through the continuity equation. \newline

The main limitation of GPR is its severe scaling with the number of data points. Standard GPR for a data set with $N$ measurements requires the inversion of the $N\times N$ covariance matrix, which has a computational cost of $O(N^3)$ and a memory storage requirement of $O(N^2)$ \citep{Rasmussen2005-dl}. There are many approximation schemes for handling the large-$N$ problem \citep{Rasmussen2005-dl, Snelson, titsias09a, Rahimi, Wilson, Williams}. In their application of GPR to astrometric data, \citet{Nelson2022} used Sparse GPR in which the likelihood function is estimated from $M\ll N$ inducing points, with a computational complexity of $O(MN^2)$. Even so, sparse GPR could not handle the GDR2 data, where $N>10^6$. For this reason, \citet{Nelson2022} took the additional step of binning the GDR2 data. Their analysis used the values of $\overline{\bs{v}}$ in 20k bins as the data and $O(10^3)$ inducing points. \newline

The usual assumption in GPR is that measurement uncertainties are independent and identically distributed Gaussian variables. When the uncertainties for individual observations are known, they can be incorporated using a heteroskedastic kernel function \citep{Rasmussen2005-dl}. In stellar dynamics, however, an additional contribution to the velocity scatter arises from the intrinsic velocity dispersion of the DF. In this work, we assume that the velocity distribution at every position in the Galaxy is a tri-variate Gaussian and that the elements of the dispersion tensor vary smoothly with position in the Galaxy. Thus, the latent function for the dispersion tensor is six dimensional. \newline

The novelty of our scheme is that the method we deploy to address the large-$N$ problem, Stochastic Variational Gaussian Process Regression (SVGPR), provides a natural path for modeling the dispersion as its own GP. SVGPR is a heavy approximation scheme that combines the framework of Variational Inference (VI) and Sparse GP methods \citep{Hensman2013, pmlr-v38-hensman15}. The key difference between SVGPR and standard GPR is that the former replaces the loss function with a lower bound approximation referred to as the evidence lower bound (ELBO). In addition, the posterior distribution over the latent process is replaced by a variational distribution. The main requirement for SVGPR is that the ELBO factorizes across individual observations. This property relaxes the assumption that the noise on measurements must be independent and Gaussian distributed, while also facilitating the application of stochastic gradient descent (SGD) as an optimization scheme. Our work extends standard SVGPR by directly exploiting this factorization property to additionally model the noise on observations, and hence the velocity dispersion, as its own GP in addition to the GP for the velocity field. \newline

In general, our method allows us to infer $\overline{\bs{v}}$ and $\bs{\sigma}$ as functions of ${\bf x}$. Since this work focuses on the development of the GPR algorithm, we restrict ourselves to the problem of inferring ${\bs{v}}$ and the diagonal components of $\bs{\sigma}$ as functions of the position relative to the mid-plane, $z$. The problem then separates into three independent analyses for, $(v_R,\sigma_R)$, $(v_\phi, \sigma_\phi)$ and $(v_z, \sigma_z)$, where $\left (R,\phi,z\right)$ are the usual Galactocentric cylindrical coordinates. In addition, since we are considering only the $z$-dependence, we restrict the data set to stars within an annular wedge centered on the Sun (see section \ref{sec:gaiadata}). This reduces the number of measurements in our analysis to $N \sim 8\times 10^5$. \newline

Our article is organized as follows. We first present an overview of Gaussian Processes, Gaussian Process Regression, and the scalable rendition we employ in Section \ref{sec:svgpregression}. We also explain our extension to the framework, and move on to discuss how we apply our scheme in the context of the general statistical model governing MW velocity observations in section \ref{sec:mostgeneralmodel}. In section \ref{sec:gaiadata}, we discuss our sample from the Gaia catalog that was used in the construction of mock data and for our inference of the vertical kinematic profile of solar neighborhood stars. We then present our choice for the GP models placed on $\overline{\bs{v}}$ and $\bs{\sigma}$, along with a summary of the algorithm and inference procedure in section \ref{sec:modelandinference}. Section \ref{sec:mockdatatests} shows the results of our tests on mock data, and our results showing smooth maps of the mean velocity and dispersion obtained from our method are summarized in Section \ref{sec:results}. In Section \ref{sec:discussion}, we discuss some of the limitations of the method's current rendition as well as possible extensions. Our results are briefly summarized, and some concluding remarks are made in Section \ref{sec:conclusion}.

%%%%%%%%%%%%%%%%%%%%%%%%%%%%%%%%%%%%%%%%%%%%%%%%%%%%%%%%%%%%%%%%
\section{SVGPR Regression} \label{sec:svgpregression}

\subsection{Overview of Regression}

Consider a set of measurements $\{y_n, x_n\}, \ n = 1, ... , N$, which we take to be noisy draws from an underlying latent function $f(x)$. The statistical relationship between the latent function and the observations is given by

\begin{equation}
\label{eq:simplegpmodel}
    y_n = f(x_n) + \varepsilon_n , 
\end{equation}

\noindent where the random variable $\varepsilon_n \sim \mathcal{N}(0, \sigma^2)$ represents independently and identically distributed Gaussian noise on the $n$'th measurement. In standard regression, we choose a parametric model $f(x; \bs{\theta})$ for the latent function and estimate the parameter values $\bs{\theta}$ such that the chosen model gives the best fit to the data. In Bayesian statistics, we specify both a prior distribution $p(\bs{\theta})$ over the model parameters and a likelihood function $p(\bs{y} \ | \ \bs{f})$ that encapsulates the relationship of equation \ref{eq:simplegpmodel} 

\begin{equation}
\label{eq:regressionlikelihood}
    p(\bs{y} \ | \ \bs{f}) \sim \mathcal{N}(\bs{y} \ | \ \bs{f}, \ \sigma^2 \bs{I}) \ .
\end{equation}

\noindent where $\bs{y} = [y_1, \ ... \ ,y_n]^T$, $\bs{f} = [f(x_1; \bs{\theta}), \ ... \ , f(x_n; \bs{\theta})]^T$, and $\bs{I}$ is the $N \times N$ identity matrix. With these quantities specified, the posterior distribution over the model parameters (up to a normalization factor) is $p(\bs{\theta} \ | \ \bs{y}) \propto p(\bs{y} \ | \ \bs{f}) p(\bs{\theta})$. Inference for the latent function proceeds by optimizing $\text{log} p(\bs{y} \ | \ \bs{f})$ with respect to $\bs{\theta}$, yielding an estimate for the best fit parameters $\hat{\bs{\theta}}$. The prediction for the output of the latent function at any novel point $x_*$ is then determined by evaluating the model $f(x_*; \hat{\bs{\theta}})$ directly.

\subsection{Overview of GPR}
\label{sec:gpoverview}

An alternative to ordinary regression is to treat the values of $f(x)$ at both observed and unobserved coordinates as components of a random vector ${\bf f}$ with a prior $p({\bf f})$. In GPR, this prior is specified by a mean function $m(x) = \mathbb{E}[f(x)]$ and a kernel function $k(x, x^\prime) \equiv \mathbb{E} [(f(x) - m(x))(f(x^\prime) - m(x^\prime))]$ \citep{Rasmussen2005-dl}. Any finite collection of latent function values is then governed by a multivariate Gaussian distribution $p(\bs{f}) \sim \mathcal{N} (\bs{m}, \bs{K})$ whose mean vector $\bs{m}$ and covariance matrix $\bs{K}$ are constructed by evaluating $m(x)$ and $k(x, x^\prime)$ at the corresponding coordinate values. We then say that $f(x) \sim \mathcal{GP}(m(x), k(x, x^\prime))$. Explicitly, the prior on the values of $f(x)$ at the $N$ observed coordinates and a finite collection of $L$ unobserved coordinates is

\begin{equation}
\label{eq:GPRprior}
p(\boldsymbol{f}, \bs{f}_*) = \mathcal{N} \bigg (\bs{m}, \begin{bmatrix}
\bs{K} & \bs{K}_*\\
\bs{K}_*^T & \bs{K}_{**}
\end{bmatrix} \bigg) ,
\end{equation}

\noindent where $\bs{K} = k(\bs{x}, \bs{x})$, $\bs{K}_* = k(\bs{x}, \bs{x}_*)$, $\bs{K}_{**} = k(\bs{x}_*, \bs{x}_*)$, $\bs{x} = [x_1, ... x_N]^T$, and $\bs{x}_* = [x_{*,1}, ... , x_{*,L}]^T$. The prior on latent function values at the observation locations is easily obtained by marginalizing the joint prior over $\bs{f}_*$. \newline 

The functions $m(x)$ and $k(x, x^\prime)$ themselves depend on a set of parameters which control the properties of the function space modeled by $p(\bs{f})$. Together with the noise variance $\sigma$, these form the set of hyperparameters $\bs{\Theta}$ of the regression model. The marginal log likelihood is given by $p(\bs{y}) = \int(p(\bs{y} \ | \ \bs{f}) p(\bs{f}) d \bs{f}$, where $p(\bs{y} \ | \ \bs{f})$ is the same as the one provided in equation \ref{eq:regressionlikelihood}. This function is optimized with respect to the hyperparameters $\bs{\Theta}$, which yields a function space that is descriptive of the data. The prediction for the latent function at any novel point $x_*$ is obtained by computing the posterior 

\begin{equation}
\label{eq:GPposterior}
    p(\bs{f}_* \ | \ \bs{y}) \propto \int p(\bs{y} \ | \ \bs{f}) p(\bs{f}_* \ | \ \bs{f}) p(\bs{f}) d \bs{f} \ .
\end{equation}

\noindent The mean and covariance of the the posterior distribution provide both the smooth estimate and corresponding predictive uncertainty for the latent function. \newline

GPR is tractable if the likelihood function is Gaussian and if the size $N$ is $\lesssim  O(10^4)$. The interested reader can consult the excellent text of \cite{Rasmussen2005-dl} for more information on GPR, including handling models with non-Gaussian likelihoods.

\subsection{GPR With Big Data - SVGPR}
\label{sec:svgprmethod}

To achieve the scalability required to analyze large astrometric datasets, we adopt the framework of SVGPR as presented by \cite{Hensman2013, pmlr-v38-hensman15}. SVGPR combats the large-$N$ problem by combining sparse GPR and Variational Inference. Being an approximation to standard GPR, SVGPR retains the same likelihood function and prior that we introduced in the previous section. \newline

In sparse GPR, one introduces a set of $M \ll N$ inducing variables and inducing points $(\bs{u} = f(\bs{z}), \bs{z})$, which serve as a low-dimensional summarization of the observational data set. By the definition of a GP, any finite collection of latent function values, including the inducing variables, forms a multivariate Gaussian. The prior on the inducing variables is then

\begin{equation}
    p(\bs{u}) \sim \mathcal{N}(\bs{m}_u, \bs{K}_{mm}) \ ,\ p(\bs{f} | \bs{u}) \sim \mathcal{N}(\bs{K}_{nm} \bs{K}_{mm}^{-1} \bs{u}, \tilde{\bs{K}}) \ ,
\end{equation}

\noindent where $\bs{m}_u = m(\bs{z})$, $\bs{K}_{mm} = k(\bs{z}, \bs{z})$, $\bs{K}_{nm} = k(\bs{x}, \bs{z})$, and $\tilde{\bs{K}} = \bs{K}_{nn} - \bs{K}_{nm} \bs{K}_{mm}^{-1} \bs{K}_{nm}^T$. In the limit $M \rightarrow N$ and $\bs{z} \rightarrow \bs{x}$, we recover the full GP prior of equation \ref{eq:GPRprior}. \newline

Even with the addition of inducing variables, the posterior of equation \ref{eq:GPposterior} still remains intractable for large $N$. Employing VI remedies this problem by introducing the variational prior 

\begin{equation}
    p(\bs{u}) \sim \mathcal{N}(\bs{\mu}, \bs{S}) \ ,
\end{equation}

\noindent where $\bs{\mu}$ and $\bs{S}$ are a mean vector and covariance matrix whose entries are free parameters. Explicit computation of the posterior is then traded for an optimization problem for a functional of the variational prior and the true prior known as the Kullback–Leibler (KL) divergence \citep{JMLR:v14:hoffman13a} 

\begin{equation}
    \text{KL}[q(\bs{u}) || p(\bs{u})] \ .
\end{equation}

\noindent By introducing both sparse GPR and VI, the log-marginal likelihood of the regression model is replaced by a lower-bound approximation called the Evidence Lower Bound (ELBO) \citep{pmlr-v38-hensman15}

\begin{equation}
\label{eq:ELBO}
    \mathcal{L}_{\text{ELBO}} = \sum_{n = 1}^N \mathbb{E}_{q(\bs{u})}[\text{log} p(y_n \ | \ \bs{u})] - \text{KL}[q(\bs{u}) || p(\bs{u})] \ ,
\end{equation}

\noindent which incorporates both the KL divergence minimization procedure and the summary of the training data via inducing variables. Since the ELBO now plays the role of $p(\bs{y})$, we optimize it with respect to the GP hyperparameters $\bs{\Theta}$, the inducing locations $\bs{z}$, and the parameters of the variational distribution $\bs{\mu}$ and $\bs{S}$. This yields a $q(\bs{u})$ that resembles the true prior, and tunes the function space modeled by the GP to be descriptive of the data. After optimization, the approximate predictive posterior providing our regression estimate for $f(x)$ is computed using $q(\bs{u})$

\begin{equation}
\label{eq:approxposterior}
    p(\bs{f} \ | \ \bs{y}) \approx\int p(\bs{f} \ | \ \bs{u}) q(\bs{u}) d \bs{u} \ .
\end{equation} \newline

Since the likelihood function for our model is Gaussian, each of the terms in the sum of the ELBO factorizes over the $N$ data points. Furthermore, the retention of $\bs{u}$ as global variables of the model enables the evaluation of the ELBO on mini-batches of data, with size $B \ll N$ \citep{Hensman2013}. This key property of SVGPR allows the application of SGD as an optimization scheme, which reduces the computational complexity of each likelihood call from $O(N^3)$ down to $O(M^3)$. Were we to marginalize over $\bs{u}$, we would introduce dependencies between the observations $\bs{y}$ and the likelihood would no longer factorize leaving us with a computational complexity of $O(N M^2)$ per likelihood call \citep{Hensman2013, titsias09a}.

%%%%%%%%%%%%%%%%%%%%%%%%%%%%%%%%%%%%%%%%%%%%%%%%%%%%%%%%%%%%%%%

\subsection{SVGPR with Input-Dependent Noise}
\label{sec:svgpr_input_dep_noise}

In many applications, the observation noise varies as an undetermined function of the input variable. In what follows, we extend the SVGPR framework to allow for joint inference of the latent function and the input-dependent noise variance. We replace our constant noise observation model of equation \ref{eq:simplegpmodel} with

\begin{equation}
\label{eq:1dobswithinputdependentnoisee}
    y_n = f(x_n) + \varepsilon_n \ , \ \varepsilon_n \sim \mathcal{N}(0, r(x_n)) \ ,
\end{equation}

\noindent where $r(x_n) = \sigma^2(x_n) + \sigma_{n, \ \text{obs}}^2$ is the total noise variance on the nth observation, composed of an unknown input-dependent variance $\sigma^2(x)$ and fixed measurement uncertainty $\sigma_{n, \ \text{obs}}^2$. \newline

\noindent Following \cite{Goldberg}, we place a GP prior on the log of the noise variances $\bs{\beta} = [\text{log}(\sigma^2(x_1)) ... , \text{log}(\sigma^2(x_n))]^T$

\begin{equation}
    \beta(x) \sim \mathcal{GP} (m_\sigma(x), k_\sigma(x, x^\prime)) \ .
\end{equation}

\noindent Crucially, our treatment preserves the Gaussian structure of the likelihood function

\begin{equation}
    p(\bs{y} \ | \ \bs{f}) \sim \mathcal{N}(\bs{y} \ | \ \bs{f}, \bs{R}) \ , \ \bs{R} = \text{diag} (r(x_1), ... , r(x_n)) \ .
\end{equation}

\noindent In turn, the factorization property of the ELBO holds, rendering SVGPR applicable to this model. \newline

Since we now have a GP for each latent function $f(x)$ and $\beta(x)$, we introduce two sets of inducing variables ($\bs{u}_f, \bs{z}_f$), and ($\bs{u}_\beta, \bs{z}_\beta$), and two variational distributions $q(\bs{u}_f)$ and $q(\bs{u}_\beta)$. Inference proceeds by optimizing the ELBO over the hyperparameters of both GPs, both sets of inducing locations, and the parameters of the two variational distributions. The predictive distribution for the latent function at novel points is still computed via equation \ref{eq:approxposterior} (with $\bs{u} = \bs{u}_f$), and the prediction for the log of the noise variances is obtained by computing 

\begin{equation}
    p(\bs{\beta} \ | \ \bs{y}) \approx \int p(\bs{\beta}  \ | \ \bs{u}_\beta) q(\bs{u}_\beta) d \bs{u}_\beta \ .
\end{equation}

%%%%%%%%%%%%%%%%%%%%%%%%%%%%%%%%%%%%%%%%%%%%%%%%%%%%%%%%%%%%%%%

\section{Probabilistic Model of Galactic Velocity Measurements}
\label{sec:mostgeneralmodel}

Suppose we have $N$ measurements of 3D stellar velocities $\bs{v}_n$ at locations $\bs{x}_n$, where $n = (1, 2, ... ,N)$. The observation model for a single measurement is 

\begin{equation}
    \bs{v}_n = \overline{\bs{v}}(\bs{x}_n) + \bs{w}_n ,
\end{equation}

\noindent where $\overline{\bs{v}}(\bs{x}): \mathbb{R}^3 \rightarrow \mathbb{R}^3$ is the latent velocity field, and $\bs{w}$ describes spatially varying scatter on each measurement. The noise on velocity measurements is governed by both the $3 \times 3$ measurement covariance matrix $\bs{\varepsilon}_{\text{obs}}(\bs{x})$ and the
velocity dispersion tensor field $\bs{\Sigma}(\bs{x}) \equiv \sigma^2_{ij}(\bs{x})$, with

\begin{equation}
    \sigma_{ij}^2(\bs{x}) = \frac{1}{\nu(\bs{x})} \int d^3 \bs{v} (v_i - \overline{v}_i)(v_j - \overline{v}_j) f(\bs{x}, \ \bs{v}) ,
\end{equation}

\noindent where Latin indices denote Cartesian coordinates. With this definition, the random vector $\bs{w}$ at the position $\bs{x}_n$ is Gaussian distributed according to $\bs{w} _n\sim \mathcal{N}(\bs{0}, \bs{\Sigma}(\bs{x_n}) + \bs{\varepsilon}_{\text{obs}}(\bs{x}_n))$. We define the stacked vectors of velocity measurements and outputs of the latent velocity field as

\begin{equation}
    \bs{V} = \begin{bmatrix}
           \bs{v}_1 \\
           \bs{v}_2 \\
           \vdots \\
           \bs{v}_n
         \end{bmatrix} \in \mathbb{R}^{3n} ,\ \ \ \ \ \ \ \ \ 
         \bs{\overline{V}} = \begin{bmatrix}
           \bs{\overline{v}}(\bs{x}_1) \\
           \bs{\overline{v}}(\bs{x}_2) \\
           \vdots \\
           \bs{\overline{v}}(\bs{x}_n)
         \end{bmatrix} \in \mathbb{R}^{3n} .
\end{equation}

\noindent This allows us to write down the likelihood function governing 3D velocity measurements

\begin{equation}
    p(\bs{V} \ | \ \bs{\overline{V}},\ \bs{K}_{w}) \propto \text{exp} \bigg[-\frac{1}{2}(\bs{V} -  \bs{\overline{V}})^T \bs{K}_{w}^{-1}(\bs{V} -  \bs{\overline{V}}) \bigg] ,
\end{equation} 

\noindent where $\bs{K}_{w} \in \mathbb{R}^{3n \times 3n}$ is the full noise covariance matrix of the model. In the most general case, there are non-trivial correlations in the noise structure across both the different data points and the three cartesian components. The general noise covariance matrix takes the form

\begin{equation}
\label{eq:fullnoisecovariance}
    \bs{K}_{w} = \text{Cov}(\bs{w}_n, \bs{w}_m) = \begin{bmatrix}
           \bs{K}_{w}^{11} & \bs{K}_{w}^{12} & \dots & \bs{K}_{w}^{1n}\\
           \bs{K}_{w}^{21} & \bs{K}_{w}^{22} & \dots & \bs{K}_{w}^{2n} \\
           \vdots & \vdots & \ddots & \vdots \\
           \bs{K}_{2}^{n1} & \bs{K}_{2}^{n2} & \dots & \bs{K}_{2}^{nn}
         \end{bmatrix} \ ,
\end{equation}

\noindent where the diagonal blocks are given by $\bs{\Sigma}(\bs{x}) + \bs{\varepsilon}_{\text{obs}}(\bs{x})$ evaluated at the measurement coordinates $\bs{x}_n$. The off-diagonal elements may be parametrized by $3 \times 3$ matrices, which encode the correlation over the 3 spatial coordinates. \newline

In addition to the above likelihood function, we also specify GP priors on the cartesian components of $\overline{\bs{V}}(\bs{x})$ and $\bs{\Sigma}(\bs{x})$. The joint prior for the stacked latent velocity field is written as

\begin{equation}
    p(\overline{\bs{V}}) \sim \mathcal{N} (\overline{\bs{V}}| \bs{m}_v, \bs{K}_v) ,\
\end{equation}

\noindent where $\bs{m} \in \mathbb{R}^{3n}$ is the stacked vector of GP mean functions at the $n$ measurement locations, and $\bs{K}_v \in \mathbb{R}^{3n \times 3n}$ is the GP covariance matrix. This prior on the stacked latent velocity field is really 3 individual GP priors governing each cartesian component of $\overline{\bs{v}}$, which are non-trivially correlated through the off-diagonal blocks of $\bs{K}_v$. Finally, we place GP priors on the log of the 6 independent components of $\bs{\Sigma}(\bs{x})$ 

\begin{equation}
    \text{log} \ \bs{\sigma}^2_{ij} \sim \mathcal{GP} (m^\sigma_{ij}(\bs{x}), K^\sigma_{ij}(\bs{x})) ,
\end{equation}

\noindent where $m^\sigma_{ij}$ and $K^\sigma_{ij}(\bs{x})$ are the mean and kernel functions of each separate GP. \newline

As currently written, inference in the most general 3D Galactic velocity model appears intractable within the framework of SVGPR. The off-diagonal blocks in the full noise covariance matrix $\bs{K}_w$ couples velocity measurements from different stars, breaking the factorization property that allows us to employ SVGPR as formulated above. In addition, the full model  would require simultaneous inference on 9 coupled GPs (3 for each component of $\overline{\bs{v}}(\bs{x})$ and 6 for the independent components of $\bs{\Sigma}(\bs{x})$), creating a high dimensional optimization problem with unfavourable computational scalability.
%In this work, we are interested in the development of our algorithm, so we leave the problem of latent function and dispersion inference in the full model for future work. 
In this work, we adopt a set of simplifying assumptions that allow us to determine non-parametric models for the latent velocity field and diagonal components of the velocity dispersion tensor as functions of $z$, the coordinate normal to the Galactic midplane, in the Solar neighbourhood. We assume that the dispersion tensor of the Milky Way is diagonal and that there are no correlations in the dispersion and latent velocity field across the different cartesian components. Under these assumptions, the input coordinates of the model reduce to scalars, and the full noise covariance becomes diagonal

\begin{equation}
\begin{split}
    \bs{K}_{w} = \text{diag}\bigg(&\sigma_1^2(x_1), \ \dots \ , \sigma_1^2(x_n) , \\ 
    &\sigma_2^2(x_1) , \ \dots \ , \sigma_2^2(x_n), \\ 
    &\sigma_3^2(x_1), \ \dots \ , \sigma_3^2(x_n) \bigg) . \\
\end{split}
\end{equation}

\noindent The GP prior on $\overline{\bs{V}}$ decouples into 3 independent GPs in each of the cartesian components

\begin{equation}
    \overline{\bs{v}}_i(x) \sim \mathcal{GP}(m_i^v(x) , k_i^v(x, x^\prime)) \ ,
\end{equation}

\noindent and the GP priors on the diagonal components of the dispersion tensor are simply

\begin{equation}
    \sigma^2_i(x) \sim \mathcal{GP}(m^\sigma_i(x) , k^\sigma_i(x, x^\prime)) \ . 
\end{equation}

\noindent Furthermore, we use one set of mean and kernel functions to build each cartesian component of the latent velocity field, and a second for the dispersion tensor GPs (see section \ref{sec:meanandkernelfuncs}). With these assumptions, the likelihood for velocity measurements factorizes across both the spatial components and the data points

\begin{equation}
    p(y_{ni} | \overline{v}_i(x_n) , \sigma^2_i(x_n) ) \propto \prod_{i = 1}^3 \prod_{n = 1}^N \text{exp}\bigg(-\frac{1}{2} \frac{[y_{ni} - \overline{v}_i(x_n)]^2}{\sigma^2_i(x_n)} \bigg) \ ,
\end{equation}

\noindent in which case the task of inference separates into three independent one-dimensional analyses in $x, y, $ and $z$. Notice that the observation model for each component of the velocity is now identical in structure to the problem considered in section \ref{sec:svgpr_input_dep_noise}. The likelihood in each cartesian coordinate satisfies the necessary factorization property discussed in section \ref{sec:svgprmethod}, allowing us to apply our SVGPR scheme to determine non-parametric models for each component of $\overline{\bs{v}}$ and the diagonal components of $\bs{\Sigma}(\bs{x})$. We leave solving the more general problem for future work.

%%%%%%%%%%%%%%%%%%%%%%%%%%%%%%%%%%%%%%%%%5
\section{Gaia DR3 Data} \label{sec:gaiadata}

We draw data from the GDR3 RVS, which has 6D phase space measurements for more than 33 million stars \citep{Gaia_DR3, Katz2023}. We transform the Gaia kinematic observables and their errors to Galactocentric rectangular coordinates $\bs{x} = (X, Y, Z), \: \bs{v} = (U, V, W)$ using the GALPY python package \citep{Bovy2015}. This is followed by a transformation to Galactocentric cylindrical coordinates $(R,\phi,Z,v_R,v_\phi,v_z)$. We choose a left-handed system such that the angular velocity vector of the disk points toward the South Galactic Pole; at the position of the Sun, $U=-v_R$, $V=v_\phi$ and $W=v_z$. We adopt the value $R_0 = 8.277 \: \text{kpc}$ for the radius of the Solar circle \citep{Gravity2022} and $Z_\odot = 0.0208 \: \text{kpc}$ for the height of the Sun above the Galactic midplane \citep{Bennett2018}. We take the velocity of the Sun with respect to the Galactic center (in Galactic coordinates) to be $(U_\odot,V_\odot,W_\odot) = (9.3,\, 251.5,\, 8.59) \: {\rm km\,s}^{-1}$ \citep{Antoja2023}. \newline

Since the focus of this paper is on the kinematics near the Sun as a function of distance from the midplane we restrict our sample to a narrow annular column centered on the Sun specified by the geometric cut
\begin{equation}
\begin{split}
    &R_0 - 0.25 \: \text{pc} \leq R \leq R_0 + 0.25 \: \text{pc} \\
    &|z| \leq 2.5 \: {\rm kpc} \\
    &|\phi| \leq 3^\circ .\\
\end{split}
\end{equation}

For simplicity, we have derived the bulk velocity of stars regardless of their spectral type and therefore do not make any cuts on magnitude or color. We return to this point in section \ref{sec:discussion}. We do make several quality cuts and corrections. In particular, we remove stars with fractional parallax error $\sigma_{\overline{\omega}}/\overline{\omega} > 0.2$. We also remove stars where the astrometric solution is suspect (Gaia parameter {\bf RUWE} $>$ 1.4). Finally, we apply the following corrections to the radial velocity measurements, which were derived by \citet{Blomme2023} to eliminate a systematic bias in the Gaia data:
\begin{equation}
    \text{RV}_{\text{corr}} = \text{RV} - 0.02755 \text{\textbf{grvs\_mag}}^2 + 0.55863 \,\text{\textbf{grvs\_mag}} - 2.81129 ,
\end{equation}
for sources with \textbf{grvs\_mag} $\geq 11$ and \textbf{rv\_ template\_teff} $< 8500$ K, and
\begin{equation}
    \text{RV}_{\text{corr}} = \text{RV} - 7.98 + 1.135 \,\textbf{grvs\_mag} ,
\end{equation}
for those with 6 $\leq$ \textbf{grvs\_mag} $\leq 12$ and 8500 K $\leq$ \textbf{rv\_template\_teff} $\leq$ 14500 K. The geometric and quality cuts leave us with a sample of 833808 stars.

%%%%%%%%%%%%%%%%%%%%%%%%%%%%%%%%%%%%%%%%%%%
\section{Inference Pipeline and GP Model Building} \label{sec:modelandinference}

%%%%%%%%%%%%%%%%%%%%%%%%%%%%%%%%%%%%%%%%%%%
\subsection{Mean and Kernel Functions} \label{sec:meanandkernelfuncs}

Our first step is to construct GP priors for the latent velocity field and dispersion tensor. This involves choosing mean functions that encode our prior knowledge of the appropriate physics, and kernel functions that govern the smoothness and correlation structure of each process. \newline

We build our velocity GP with a mean function $m(z) = 0$. This choice aligns with the null hypothesis that the Galaxy is in dynamical equilibrium, and so is symmetric about the midplane. Under such conditions the radial and vertical components of the first velocity moment vanish \citep{Binney2011}. The azimuthal component of $\overline{\bs{v}}$, which is non-zero due to Galactic rotation, is removed in our pre-processing (see section \ref{sec:inferenceprocedure}). For the velocity dispersion, we use a non-zero mean function to reflect the known behavior of the three diagonal components of the dispersion tensor as functions of $z$. In equilibrium, the existence of thin and thick disc components gives rise to an increase in the dispersion as a function of $|z|$ \citep{Li2021}. To account for this increase, we assume the following mean function for each of the three dispersion tensor components

\begin{equation}
\label{eq:tanhfit}
    m_\sigma (z) = A + B\, \text{tanh} \bigg ( \frac{|z - z_0|}{l_0} \bigg ).
\end{equation}

\noindent The parameters $(A, B, z_0, l_0)$ are different for the $R$, $\theta$, and $z$ components of the dispersion tensor. \newline

In preliminary tests, we tried two kernels for both the mean velocity and velocity dispersion. The Radial Basis Function (RBF) kernel 

\begin{equation}
\label{eq:RBFkernel}
    K_{\overline{v}} (z, z^\prime) = \lambda^2 \text{exp} \bigg ( -\frac{|z - z^\prime|^2}{2 l^2} \bigg), 
\end{equation}

\noindent and the Rational Quadratic (RQ) kernel

\begin{equation}
    k_{\sigma} (z, z^\prime) = \lambda_\sigma^2 \bigg (1 + \frac{(z - z^\prime)^2}{2 \alpha \l_\sigma^2} \bigg )^{-\alpha} .
\end{equation}

\noindent The RBF kernel is the default covariance function for GPR and has the attractive property of being both simple and infinitely differentiable. It is governed by two hyperparameters $\bs{\Theta} \equiv (\lambda ,\ l)$, which control the amplitude and correlation length of the latent function. The RQ kernel is equivalent to adding many RBF kernels together and therefore allows for the modeling of features across many length scales. In addition to $\lambda$ and $l$, it has a third hyperparameter $\alpha$ that controls the relative weighting of small-scale and large-scale features. In the limit $\alpha \rightarrow \infty$, the RQ kernel reduces to the RBF kernel. In our preliminary tests of the method, we found little difference in performance between the RBF and RQ kernels for the mean velocity but did find that the RQ kernel performed better for the velocity dispersion. We therefore adopt the RBF kernel for the latent process and the the RQ kernel for the dispersion process in the analyses presented in the remaining sections. 

\subsection{Outline of the Inference Procedure}
\label{sec:inferenceprocedure}

\begin{figure}[h!]
\centering
\includegraphics[width=0.5\textwidth]{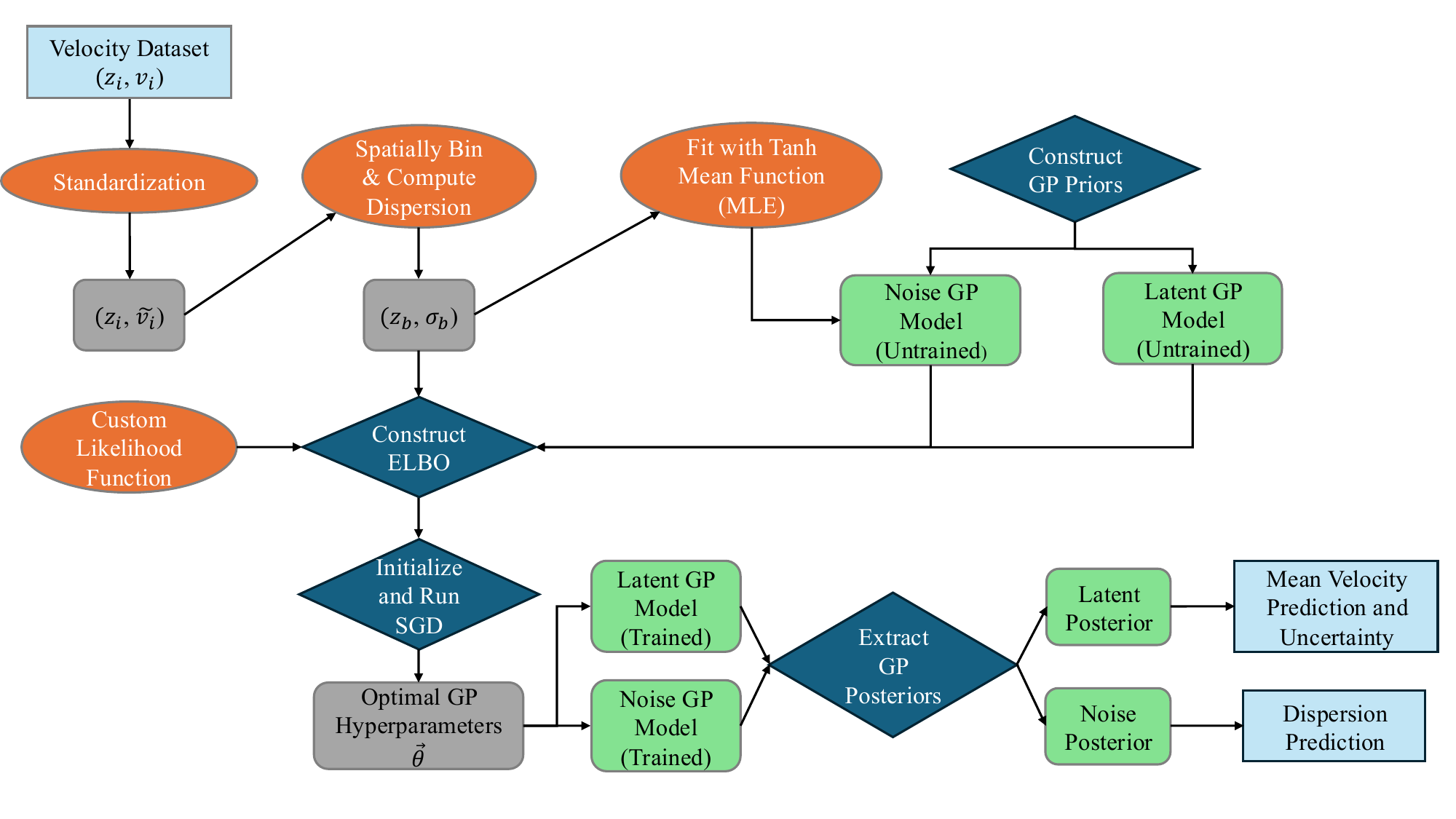}
\caption{Flowchart illustrating the steps of our SVGPR inference procedure. Light blue boxes indicate the pipeline's inputs and outputs. Orange ovals represent data preprocessing steps before passing to GPyTorch. Blue diamonds denote inference tasks executed directly within GPyTorch. Green boxes correspond to our GP models, and grey boxes indicate intermediate results.}
\label{fig:inferenceflowchart}
\end{figure}

We implement our algorithm using GPyTorch, a fast and robust toolbox for performing GPR \citep{gardner2021}. Our custom code handles the injection of a second GP into the GPyTorch inference pipeline and necessary data pre-processing. Inference in our setup proceeds according to the flow chart depicted in figure \ref{fig:inferenceflowchart}. Given a training set of a single velocity component as a function of $z$, we standardize the data by subtracting its mean and dividing by its standard deviation. The GP priors for both the latent velocity and dispersion processes are then constructed according to the prescription outlined in section \ref{sec:meanandkernelfuncs}. We proceed by spatially binning the velocity measurements and compute the velocity dispersion using conventional statistical methods. This binned dispersion profile is then fit with the mean function of equation \ref{eq:tanhfit} using maximum likelihood estimation (MLE). This mean function is held fixed throughout the subsequent regression procedure. \newline

Next, we instantiate the SGD optimizer provided by GPyTorch. The operation of SGD is controlled by an additional set of algorithmic hyperparameters, which include the optimization learning rate, the number of training steps, and the SGD batch ratio. In Section \ref{sec:mockdatatests}, we explain the tests that inform our best choice for these additional hyperparameters. The SGD optimizer is then run on the ELBO to determine the optimal set of hyperparameters for both GPs that best describe the data. Finally, we invoke GPyTorch to compute the posterior distributions for both velocity moments. \newline

The prediction and corresponding uncertainty for the mean velocity field as a function of $z$ are obtained by directly querying the parameters of the corresponding posterior distribution. In contrast, the prediction and uncertainty for the velocity dispersion tensor are estimated in a different manner since the covariance matrix of the dispersion posterior cannot be interpreted in the same way as the covariance matrix of the mean velocity. This arises from the fact that the GP prior governing the dispersion process is not placed on observations but instead governs what would be the noise hyperparameters in a heteroskedastic treatment of regression. In the absence of a rigorous approach, we estimate the uncertainty on the dispersion in a qualitative fashion. Given our initial prediction for the dispersion, we run our inference pipeline repeatedly on smaller subsets of our observational data. The rationale behind this approach is that we expect real kinematic features in the dispersion profile to persist even when the density of information in the dataset is reduced. We attribute any features that change significantly or vanish when the data set is reduced to statistical fluctuations. We perform this procedure $n$ times until we see appreciable change in the dispersion output, leaving us with a set of predictive curves for the dispersion $\{{\sigma_n}(z)\}$ that are obtained from the GP posterior. We then evaluate the mean and standard deviation of these curves

\begin{equation}
    \overline{\sigma}(z) = \frac{1}{N} \sum_{i = 1}^n \sigma_i(z) \ , \ s(z) = \sqrt{\frac{1}{n - 1} \sum_{i = 1}^n(\sigma_i(z) - \overline{\sigma}_i(z))^2} \ ,
\end{equation}

which provides us with our estimate for the dispersion profile and its corresponding confidence region. This entire procedure is illustrated when presenting our results in section \ref{sec:results}. We recognize that this approach is ad hoc, and we touch on this point in section \ref{sec:discussion}. \newline

In both our mock data tests and our analysis of the Gaia data, we work with $N \simeq 800,000$ stars. We use $M = 1000$ inducing points for both GPs and an SGD batch ratio of 100. The predictive calculation takes $\approx$ 10 hours on a single Nvidia RTX 3070 Ti GPU paired with an intel i9 10850K CPU. If the calculation was done with exact GPR, the CPU time would have increased by a factor of $N^3 / M^3 \simeq 10^6$. \newline

%%%%%%%%%%%%%%%%%%%%%%%%%%%%%%%%%%%%%%%%%%%%
\section{Mock Data Tests} \label{sec:mockdatatests}

\begin{figure}[h!]
\centering
\includegraphics[width=\columnwidth]{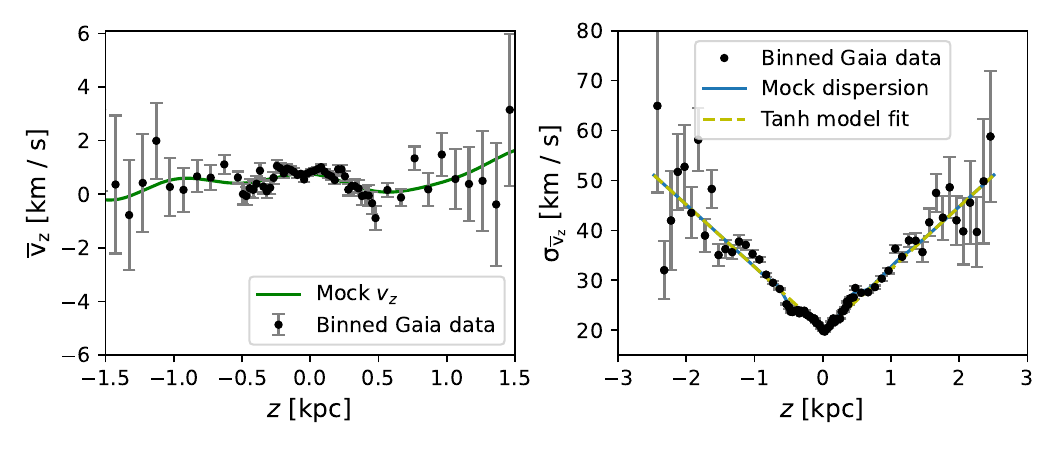}
\caption{Mock latent vertical velocity data constructed to match the qualitative features seen in the vertical velocity of solar neighbourhood stars. Left panel: Vertical mean velocity of stars obtained from binning (black dots) and our mock mean profile obtained from a simple GP fit (green curve). Right panel: Binned vertical dispersion profile (black dots) and our mock $z$-dependent dispersion built from fitting equation \ref{eq:tanhfit} to the binned data.}
\label{fig:mockvzdata}
\end{figure}

We first test our algorithm on mock data in order to establish its capability to pick out features in the data and determine appropriate algorithmic hyperparameters for the SGD optimizer. We construct mock data with properties similar to those of the real data by the following procedure. First, we bin our sample in $z$ with a bin size of $\Delta z = 25$ pc for stars lying within $0.5$ kpc of the mid-plane, and $\Delta z = 100$ pc for stars beyond $|z| = 0.5$ kpc. This prescription combats the exponentially decreasing number density as one moves away from the Galactic midplane. In these bins, we compute $\overline{v}_z$ and $\sigma_{\overline{v}_z}$ using conventional techniques. Figure \ref{fig:mockvzdata} shows the resulting binned data (black dots with error bars). Next we run a simple GPR analysis on the mean velocity profile with fixed $(l,\lambda) = (3 \ \text{km/s}, \ 0.4 \ \text{kpc})$. The result is shown by the green curve in the left panel of figure \ref{fig:mockvzdata}. For the mock dispersion profile, we first fit the dispersion profile to equation \ref{eq:tanhfit} (yellow dashed curve). This captures the large-scale behavior due to the transition from thin to thick disk as one moves away from the midplane. We then add Gaussian features at $z\simeq \pm 400\,{\rm pc}$ to mimic the features seen in the profile. The result (blue curve) has the general features of the real data.

 The standard procedure for determining suitable run-time parameters in a machine learning model is to conduct a grid search \citep{Liashchynskyi}. The learning process is carried out for each permutation of parameter to determine the combination that best facilitates convergence. In SVGPR, the relevant parameters are the number of steps of gradient descent, i.e., training steps, the learning rates for both the latent and dispersion GPs, the batch ratio for SGD, and the number of inducing points used to summarize both the mean velocity and dispersion processes. The learning rates of both GPs separately dictate how large of a step in hyperparameter space the optimizer takes in every gradient update. We decouple the learning rates for the two GPs to allow for more flexibility in the algorithm. These parameters do not affect the computational complexity of inference. The batch ratio $B_R \equiv \frac{N_{\text{tot}}}{N_{\text{batch}}}$ sets the size of a mini-batch of data that SGD operates on during each training step. Varying this parameter alters both the computation time of the algorithm and the quality of inference, as a mini-batch serves as one unit of information that the ELBO operates on. Finally, the number of inducing points $M$, which, in principle, dictates how well the salient information in the training data is summarized, yields the largest contribution to the computational complexity of inference ($O(M^3)$). We construct our grid of parameters using the values in table \ref{tab:paramgrid} in Appendix \ref{app:gridsearchresults}. \newline

Since $M$ has the largest impact on the computation time, we hold its value fixed at $M = 100$ when performing our grid search. Once the optimal batch ratio, learning rates, and number of training steps are determined, we can increase $M$, which is expected to systematically improve the fit quality. In this section, we present fits to our mock data when adopting the best combination of algorithmic parameters from our grid search. Detailed results from the search can be found in Appendix \ref{app:gridsearchresults}. \newline

Figure \ref{fig:bestmockdatafit} shows the results of applying our SVGPR inference pipeline to the mock data presented in figure \ref{fig:mockvzdata}. The top row of plots depict the GP fit to the latent velocity while the bottom row shows the prediction for the dispersion profile. The optimizer is able to discern the correct amplitude for the latent velocity field, but fails to extract the small scale fluctuations that appear on scales $\lesssim O(1)$ km/s. Close to the midplane, the latent fit residuals (calculated as $(\overline{v}_z - \overline{v}_z^{\text{true}}) / \sigma$, where $\sigma$ is the standard deviation of our full dataset) are minimized, which is explained by the fact that the density of data points is highest in this region. For the dispersion process, the optimizer correctly extracts the entire profile, with minimal residuals ($\sigma_{\overline{v}_z} - \sigma_{\overline{v}_z}^\text{true}$). Just as in the case for the latent velocity, the residuals get worse with increasing $|z|$ due to the exponential suppression in the density of data points. At this stage, there is a clear disparity in SVGPR's ability to fit both processes to the same degree of precision. A reasonable assumption is that increasing the number of inducing points may remedy this problem. \newline

\begin{figure}[h!]
\centering
\includegraphics[width=\columnwidth]{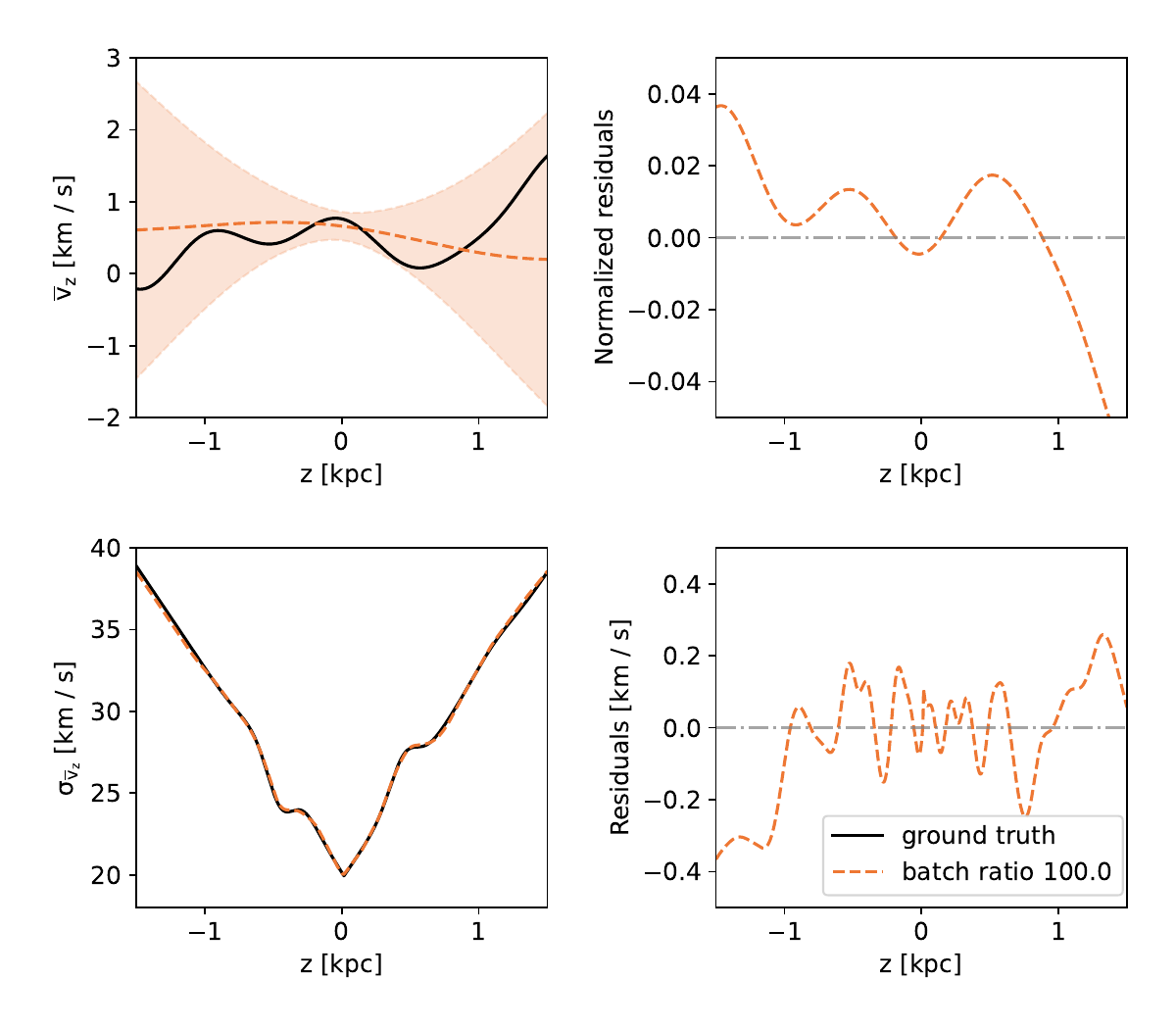}
\caption{SVGPR inference results on our mock data set, performed using the optimal set of algorithmic parameters as dictated by our grid search (summarized in table \ref{tab:optimalinferenceparameters}, except here we use M = 100 as discussed above). Top left panel: Regression prediction for the mean vertical velocity field. The prediction (orange dashed-curve) and corresponding 95 \%  confidence region (solid orange band) are obtained from the latent velocity posterior GP. Top right panel: Residuals of the predicted profile from the ground truth (solid black curve in the top left panel), normalized by the standard deviation of the mock data set. Bottom left panel: SVGPR prediction for the vertical velocity dispersion. Bottom right panel: Residuals of the predicted dispersion profile and the ground truth (black solid curve in the bottom left panel).}
\label{fig:bestmockdatafit}
\end{figure}

\begin{figure}[h!]
\centering
\includegraphics[width=\columnwidth]{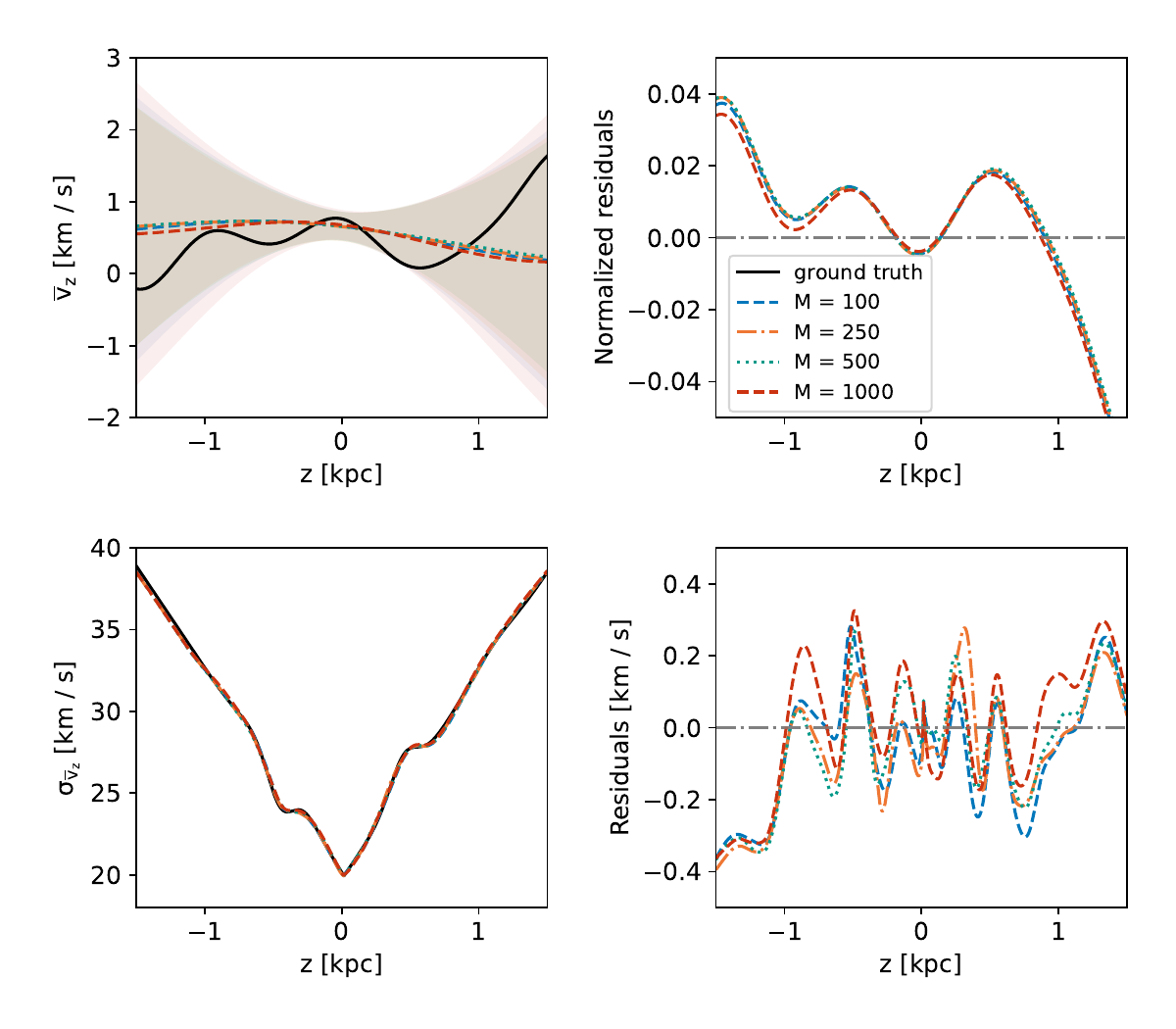}
\caption{Same as figure \ref{fig:bestmockdatafit}, where we now vary the number of inducing points for both GPs. The remaining algorithmic parameters were held fixed at the values provided in table \ref{tab:optimalinferenceparameters}.}
\label{fig:inducingpointstestvz}
\end{figure}

\begin{figure}[h!]
\centering
\includegraphics[width=\columnwidth]{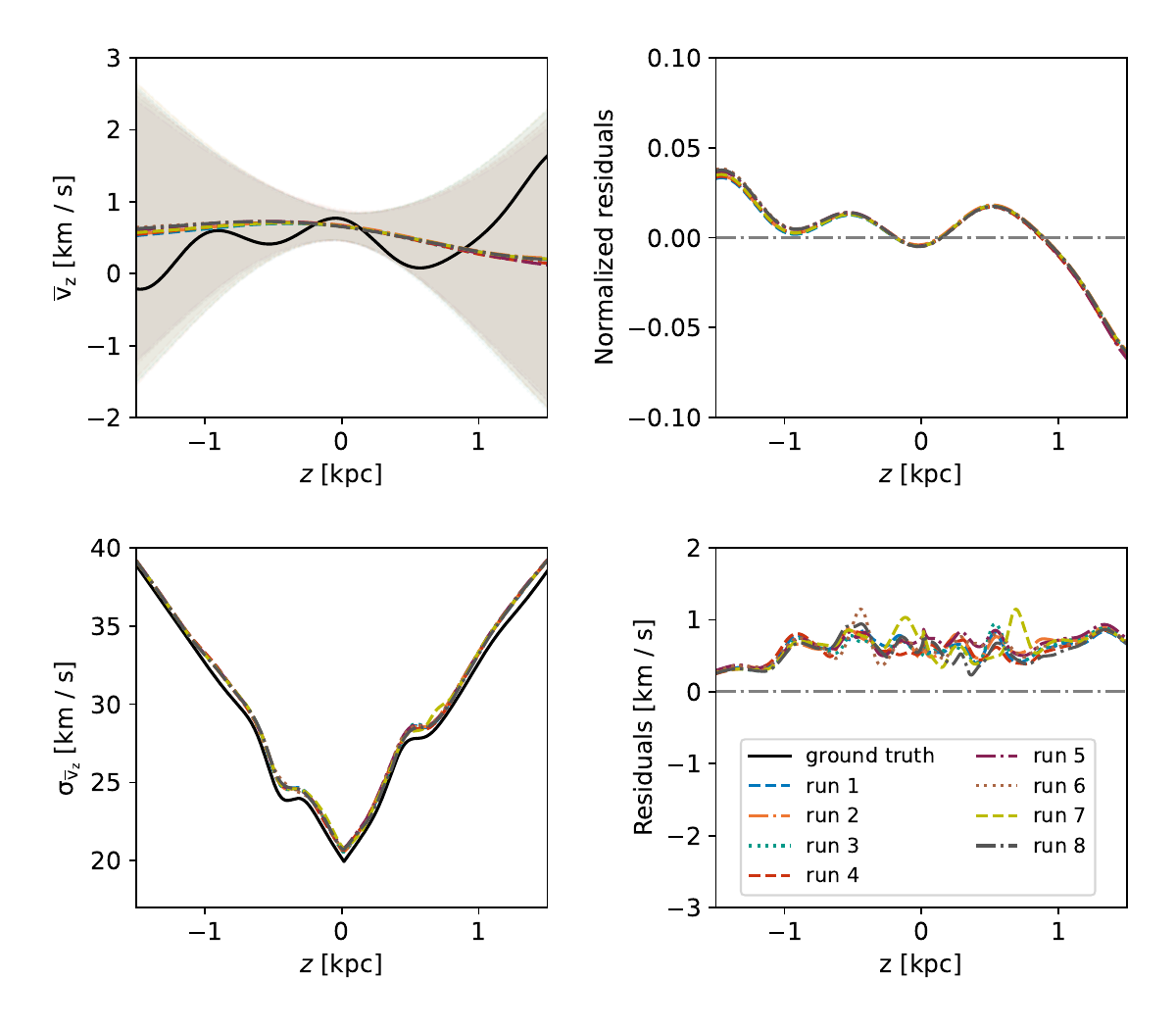}
\caption{Same as figure \ref{fig:bestmockdatafit}, where we now show the resulting predictions for 8 different random mini-batch selections by the SGD optimizer. Once again, all algorithmic parameters were kept fixed at the values presented in table \ref{tab:optimalinferenceparameters}.}
\label{fig:gpseedtest}
\end{figure}

Figure \ref{fig:inducingpointstestvz} shows the results of our test where we study the effect of increasing $M$ with all other algorithmic parameters kept fixed. The fit variation we observe with $M$ is not nearly as large as expected. To understand whether these changes are truly due to variation in the number of inducing points or simply a reflection of the inherent stochasticity of the SGD optimization process, we perform an additional test. We fix all algorithmic parameters including $M$ and perform 8 inference runs, where the only difference is the random selection of mini-batches of data by the optimizer. The resulting fits (shown in figure \ref{fig:gpseedtest}), demonstrate that the variability between different mini-batch selections are comparable to, and sometimes exceed, the fit variation we observe when changing the number of inducing points. This suggests that the small changes we see in figure \ref{fig:inducingpointstestvz} may be attributed to statistical fluctuations inherent to the optimization process. \newline

Given this finding, the direct interpretation of figure \ref{fig:inducingpointstestvz} is that the information in our training data is adequately summarized with only 100 inducing points. This makes sense for the dispersion process, as the bulk behavior is already captured by the fitted tanh mean profile. From the standpoint of the latent process however, this is puzzling. Adequate summary of the latent process with $M / N \sim 10^{-4}$ is in friction with the standard behavior expected of SVGPR with inducing points \cite{Hensman2013}. Even though the role of the inducing points is better understood for the case where one only fits a GP for the latent function, it is not clear why this would break down in moving to the regime where the noise is also modeled by a GP. An important result from our tests shown in figure \ref{fig:gpseedtest} is the systematic offset exhibited by the dispersion fits. Despite holding all algorithmic parameters at the same values as the tests presented in figures \ref{fig:bestmockdatafit} and \ref{fig:inducingpointstestvz}, this offset appears only in our stochastic runs. This suggests that the joint optimization landscape of the ELBO is highly populated by local minima that each describe statistically equivalent fits to the data but differ in their absolute scale, with the dispersion process appearing particularly susceptible to this. \newline

We hypothesize that the apparent disparity in the ability to fit both the latent and dispersion processes to the same level of precision has little do with $M$, and is instead an issue of scales. The fluctuations we have built into our mock latent processes manifest on velocity scales of $O(1)$ km/s, whereas the overall amplitude of $\sigma_{v_z}$ is an order of magnitude larger ($\sim 30$ km/s). Features in the latent velocity field are effectively buried in noise, rendering it difficult to extract information about the latent process even in the limit when $M \rightarrow N$. This scale disparity would also explain the difference in the outcomes of the dispersion fits between figures \ref{fig:bestmockdatafit} and \ref{fig:gpseedtest}. The larger scale of the dispersion process may be mapping small changes in the optimization trajectory to noticeable absolute offsets, and the joint nature of optimization would allow for trade-offs where slightly different dispersion profiles can compensate for small variations in the latent velocity fit. \newline

To test if this is truly a scale issue, we regenerate our mock data with the same mock profiles as in figure \ref{fig:mockvzdata}, but artificially raise the magnitude of the mean vertical velocity by a factor of 30. We then execute our inference pipeline using the same set of algorithmic parameters, including $M = 100$. The results of this test are shown in figure \ref{fig:inflatedfit} below.

\begin{figure}[H]
\centering
\includegraphics[width=\columnwidth]{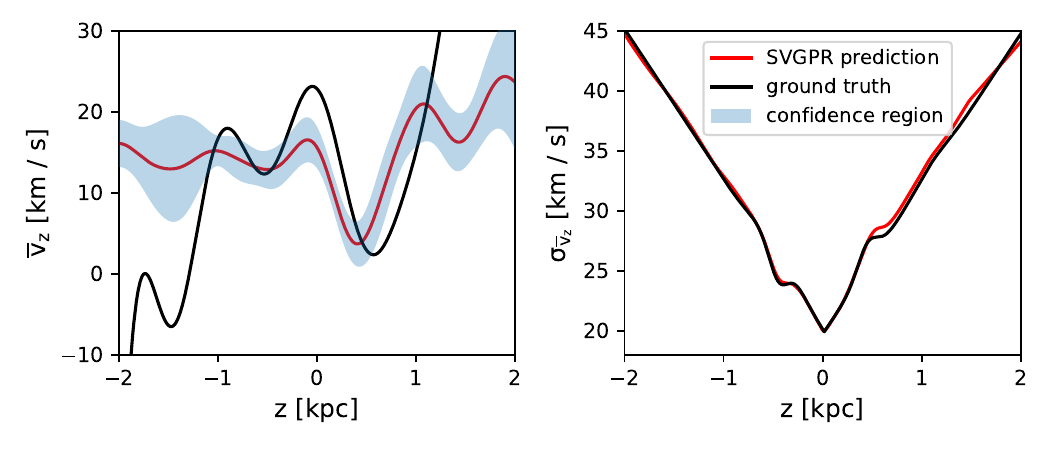}
\caption{SVGPR predictions for the latent vertical velocity and velocity dispersion for our artificially inflated mock data. Left panel: regression prediction for the mean vertical velocity (red solid curve) and corresponding 95 \% confidence region (blue band). Right panel: SVGPR prediction for the dispersion profile (solid red curve). The ground truth for each process are provided by the solid black curves.}
\label{fig:inflatedfit}
\end{figure}

The fit performance for the latent process is clearly improved in this case. The optimizer has picked out the right length scale for fluctuations, and performs better at fitting mean $v_z$ near the origin. We also see a slight decrease in the fit quality for the dispersion process. Since the latent and dispersion process are much closer in scale, we are seeing the optimizer strike a balance when fitting both processes. The results of this test support our hypothesis that the apparent lack of improvement in regression quality with increasing $M$ is not indicative of optimal inducing point summarization, but is reflective of the challenge of attempting to jointly optimize two GPs modeling processes with largely different amplitudes. This is made especially clear when noting the asymmetry in the fit quality with respect to $z$. Below the mid-plane, the dispersion amplitude is roughly $5$ km/s smaller than above the mid-plane. Consequently, the latent fit performs better for $z < 0$, where the ratio of signal-to-noise is smaller. While the fit quality in this test is not ideal, if this were a real data set one would perform a full grid search (including different values of $M$) to aid in convergence. This also demonstrates that every data set presents a unique optimization problem. \newline

We conclude that our method's strength is its ability to determine the detailed structure of the velocity dispersion profile. While the method performs worse at learning fine-grained structure in $\overline{v}_z$, it still provides an adequate estimate for its large-scale behavior. We have demonstrated that this is not due to a problem with the method, and instead is a unique challenge presented by the particular data we wish to analyze. Table \ref{tab:optimalinferenceparameters} summarizes the optimal set of optimization parameters that should be employed when performing inference on real vertical velocity data from Gaia. Finally, we note that despite seeing no improvement with larger $M$, we choose to adopt $M = 1000$ in analyzing our real data. Although qualitatively similar, our mock data and the real Gaia data are not the same. In the event that features in the dispersion appear on smaller length-scales than what we have baked into our mock tests, we wish to have the ability to adequately extract them. 

\begin{table}[htbp]
    \centering
    \caption{Summary of the optimal algorithmic parameters from our grid search results discussed in appendix \ref{app:gridsearchresults}.}
    \label{tab:optimalinferenceparameters}
    \begin{tabular}{|c|c|}
        \hline
        \textbf{Parameter} & \textbf{Value}  \\
        \hline
        Training Steps & 300  \\
        \hline
        Batch Ratio & 100.0  \\
        \hline
        Latent Learning Rate & 1.0   \\
        \hline
        Dispersion Learning Rate & 0.1  \\
        \hline
        Inducing Points & 1000   \\
        \hline
    \end{tabular}
\end{table}

%%%%%%%%%%%%%%%%%%%%%%%%%%%%%%%%%%%%%%%%%%%%%%%%%%%%%%%%%%%%%%%%%

\section{Results} \label{sec:results}

In this section we present the results of our SVGPR analysis of our GDR3 sample. Hyperparameter learning was performed with 300 training steps, a learning rate of 1.0 and 0.1 for the latent and dispersion processes, a batch ratio of 100, and $M = 1000$ inducing points for both GPs. Before discussing our inferred velocity  profiles, we demonstrate our procedure for obtaining the dispersion prediction and its error as discussed at the end of section \ref{sec:inferenceprocedure}. \newline

\begin{figure}[h!]
\centering
\includegraphics[width=\columnwidth]{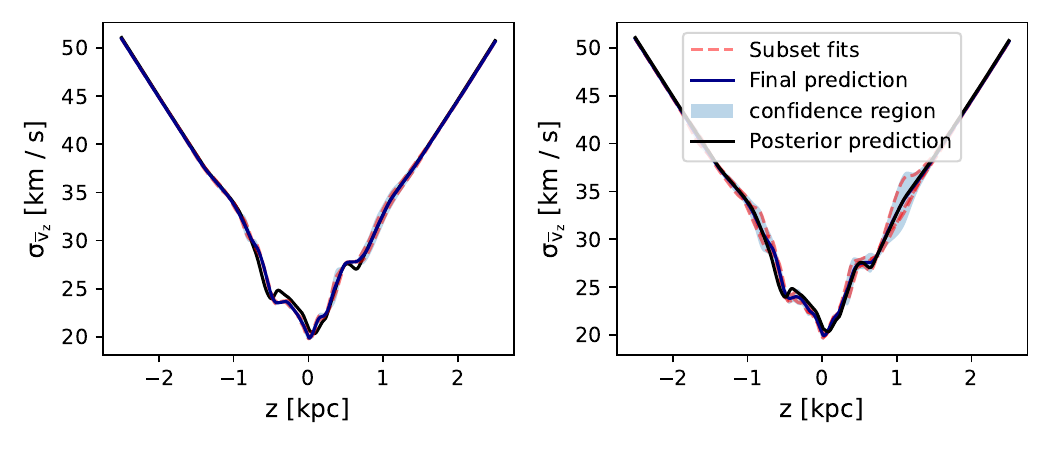}
\caption{Illustration of our scheme for estimating the statistical uncertainty on our learned dispersion profiles. Left panel: SVGPR fits for the dispersion for two subsets of the total training set (red-dashed curves). Each subset corresponds to exactly half of the training set. We also show the regression output when applied to the entirety of the data. Right panel: Same as in the left panel, this time trained on four equal sized segments of the training set.}
\label{fig:DispErrorEst}
\end{figure}

Figure \ref{fig:DispErrorEst} shows the dispersion prediction obtained from our SVGPR pipeline when applied to subsets of the total velocity dataset. The black curve shown in both panels is the regression prediction obtained as the mean of the dispersion GP posterior when our inference pipeline is applied to the entire dataset. We then split our data into two halves (left panel), each containing every odd and even data point respectively, and run our SVGPR pipeline on both. The sets of red dashed curves in the left panel are then the resulting predictions extracted from the GP posterior from both of these runs. We then continue to split our data into four equal halves and run each through our regression pipeline. The red dashed curves in the right panel show the resulting dispersion predictions for these 4 substets of the data. Taking the mean of the red-dashed curves in both cases gives us a final prediction for the vertical dispersion, shown as the solid blue curves. The standard deviation of the sets of red-dashed curves are then used to compute the 95 \% confidence interval on our final dispersion prediction. As can be seen in the left panel, some features present in the fit to the full dataset begin to either fluctuate strongly or disappear when the analysis is restricted to data subsets (at $z \approx - 0.6$ kpc and $z \approx 1$ kpc). Based on the criteria discussed in section \ref{sec:inferenceprocedure}, we derive our final dispersion prediction and uncertainty from the left panel of figure \ref{fig:DispErrorEst}. This procedure was performed identically for the azimuthal and radial components of the dispersion. \newline

\begin{figure}[h!]
\centering
\includegraphics[width=\columnwidth]{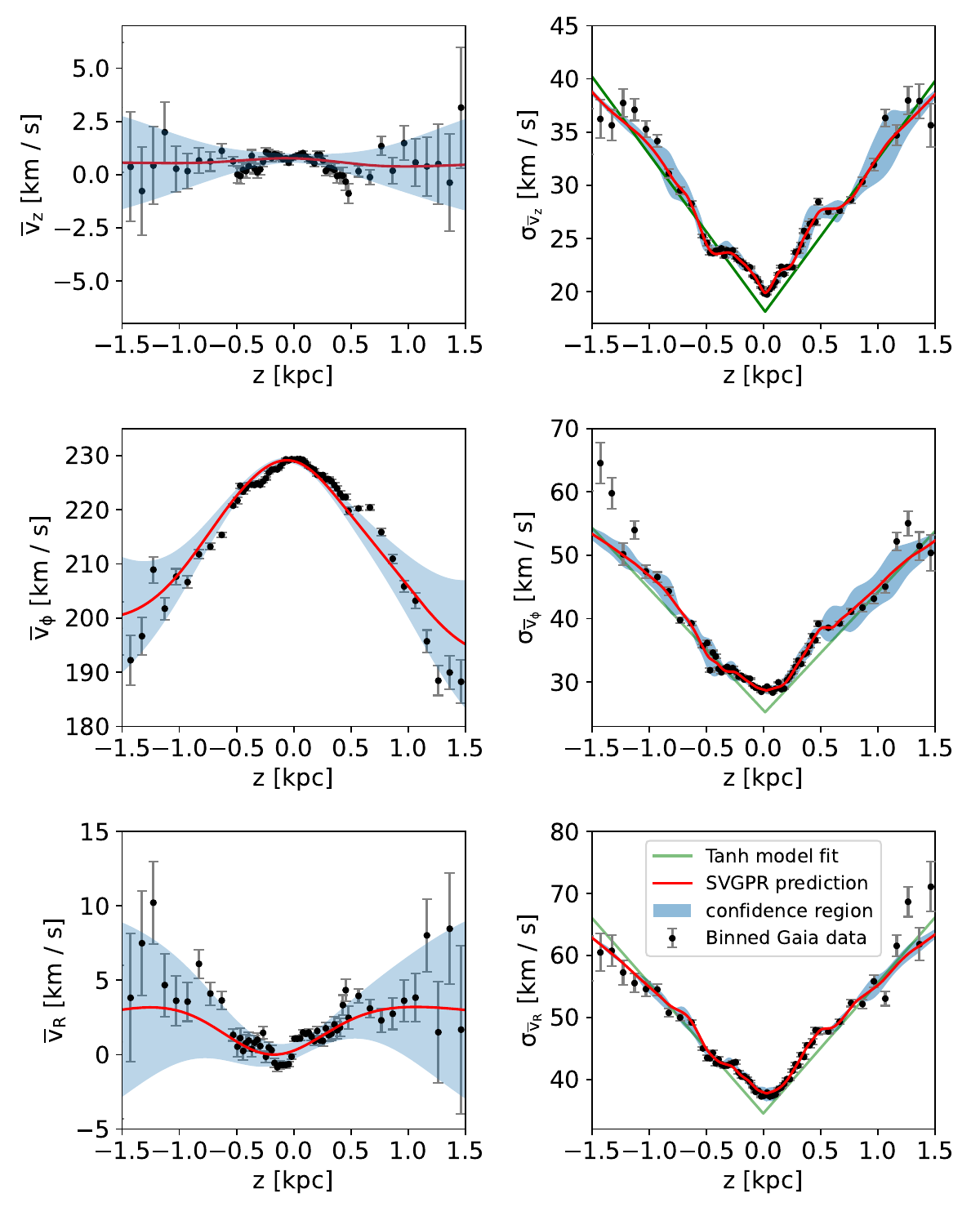}
\caption{Our SVGPR predictions for the three components of the latent velocity field and velocity dispersion tensor as functions of Galactic $z$. In all panels, the black dots show the kinematic profiles obtained from binning our stellar sample in $z$, and the red solid curves depict the predictions of our SVGPR scheme. The blue shaded bands give the 95 \% confidence intervals on the regression predictions. Top left panel: Inferred mean vertical velocity profile and corresponding uncertainty. Middle left panel: Inferred azimuthal mean velocity and corresponding uncertainty. Bottom left panel: Inferred radial velocity and corresponding uncertainty. Top right panel: Inferred vertical velocity dispersion profile and corresponding uncertainty. Middle right panel: Inferred azimuthal velocity dispersion and corresponding uncertainty. Bottom right panel: Inferred radial velocity dispersion and corresponding uncertainty. For all three plots in the right column, we also provide the tanh fits to each binned profile used in the construction of the dispersion GP's mean function (green solid curves).}
\label{fig:SVGPRpredictions}
\end{figure}

Figure \ref{fig:SVGPRpredictions} show our results for the latent velocity field and velocity dispersion as functions of $z$. In each panel, the black data points show the velocity moments obtained from binning the sample. The red curves in the left column are the model predictions for each component of the mean velocity field, and the same coloured curves in the right column give the dispersion profiles, obtained from the procedure demonstrated in figure \ref{fig:DispErrorEst}. The blue bands give the confidence regions on both sets of predictions. In the case of the latent velocity field, the predictions and confidence regions are obtained directly from the latent GP posterior. \newline

Focusing on the latent velocity fits in the regions beyond $|z| \sim 1.5$ kpc, we see the SVGPR predictions rebound upwards towards the mean value that the latent velocity field has near the midplane of the disc. This behavior is caused by the interplay between standardizing our data for performing inference and the exponential drop off in the density of data points with $|z|$. The GP prior for each velocity component is built with a mean function of 0, and in regions with low data count, the regression output tends back to this prior. When the standardization is reversed after inference, the true mean originally present in the dataset is restored, and the resulting regression prediction at large $|z|$ converges towards the true mean. In general, we must be careful in interpreting our fits in regions where the confidence regions begin to blow up. Comparing the smooth prediction for the azimuthal and radial mean velocity components to the corresponding binned profiles, we see good agreement in their large scale behaviour. Comparison of our smooth latent velocity maps with the results of \cite{Nelson2022} show reasonable agreement on a qualitative level, with our inferred $\overline{v}_z$ profile differing the most. In both cases, the velocity field dips on either side of the galactic disc, but in our case we find a decrease in $\overline{v}_z$ of less than $\sim 0.5$ km/s whereas \cite{Nelson2022} report close to double this value. In addition, the decrease in mean vertical velocity appears at roughly $|z| \simeq 1$ kpc in contrast with $|z| \simeq 0.5$ kpc in the maps of\cite{Nelson2022}. The most likely cause for this discrepancy is the different approach to GPR that we take here, being a much more aggressive approximation to standard GPR than sparse GP alone. In addition, they performed their GP fits to data that was first binned, whereas we apply our method to the direct velocity measurements. A component of this discrepancy may also be due to selection effects, which we touch on in section \ref{sec:discussion}. \newline

Turning to the right panel of figure \ref{fig:SVGPRpredictions}, our dispersion predictions show interesting features on small scales for all three components. In every case, the structure of fluctuations are asymmetric about the midplane of the disc. We take the stance that these features are physical given the results of our mock tests, and interpret them as a sign that the vertical dynamics of the disc are in a state of disequilibrium. Upon closer inspection, in the region close to the midplane of the disc, the features in all three inferred dispersion profiles occur at roughly the same locations in $z$, despite the difference in scale of each profile. Every profile appears to have a strong dip located at $z \sim -0.5$ kpc that is matched with an enhancement above the disc at $0.5$ kpc. \newline

\subsection{Features in the Dispersion Maps}
\label{sec:dispersionfeatures}

\begin{figure}[h!]
\centering
\includegraphics[width=\columnwidth]{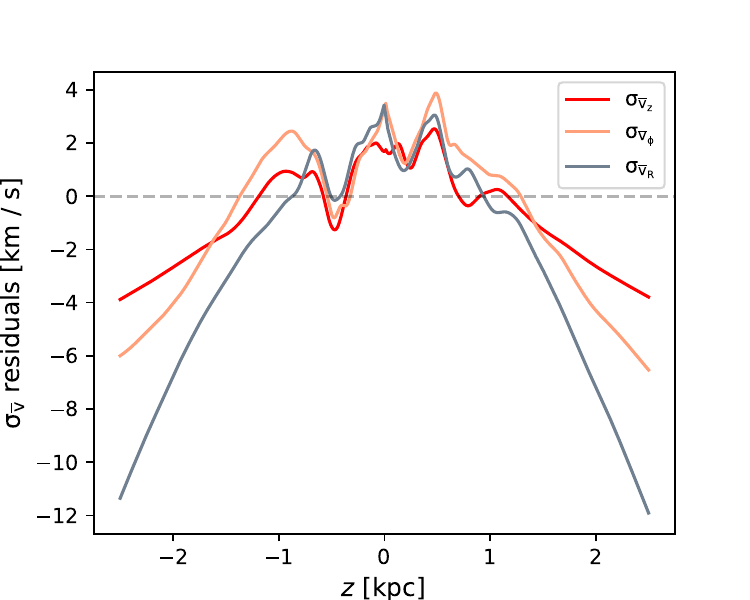}
\caption{Residuals of each of our dispersion profile fits with respect to the tanh parametric fits given by the green curves in the right column of figure \ref{fig:SVGPRpredictions}. The grey-dashed line helps guide the eye to zero.}
\label{fig:dispersionpredresiduals}
\end{figure}

While the purpose of this work is primarily to present our method for inferring smooth velocity moments, it is interesting to inspect the dispersion features more closely. We do so by computing the residuals of the SVGPR dispersion predictions with respect to their tanh parametric fits that were utilized for GP construction (the green solid curves in the right panel of figure \ref{fig:SVGPRpredictions}). Figure \ref{fig:dispersionpredresiduals} depicts the resulting residuals. Paying attention to the region within $\pm 1$ kpc of the midplane of the disc, all of the features appear to possess a degree of correlation across $\sigma_{v_z}, \: \sigma_{v_\phi}, \:$ and $\sigma_{v_R}$. The the strong dip at $z \sim -0.5$ kpc is echoed across all of the dispersion profiles. At the same $|z|$ above the midplane, we see prominent enhancement, which reflects the features seen in the fits of figure \ref{fig:SVGPRpredictions}. The fluctuations in very close proximity to the midplane near $|z| \sim 0.1$ kpc also exhibit a strong degree of similarity across each of the dispersion maps. Since these features are ubiquitous across all three profiles, it is interesting to consider the possibility that they all originate from a common process. \cite{Guo2022} found a correlation between the locations where arms of the Gaia $z - v_z$ phase spiral intersect the $z$ axis and suppression/enhancement of the vertical velocity dispersion. Interestingly, they find that the phase spiral intersects the $z$ axis in phase space around $|z| \sim 0.5$ kpc \citep{Antoja2023, Guo2022} which is where we see similar suppression/enhancement in our inferred dispersion maps.

\section{Discussion} \label{sec:discussion}

The main goal of this work is to build mathematical models of the mean velocity and velocity dispersion tensor that would allow for the systematic investigation of the dynamical properties of the Galaxy through Jeans analysis. The strength of our method lies in its ability to construct differentiable kinematic maps of the Galaxy with minimal input beyond observations of the 6D phase space coordinates of stars. Furthermore, since the method is non-parametric, it has utility in studying departures from equilibrium that are imprinted on the kinematic structure of stars. Though our choices in the constructing GP priors were made to reflect our knowledge thus far on the properties of the first and second velocity moments of the DF, they are not in any way restrictive. The kernel functions of both GPs are perfectly capable of adapting to features that may originate from effects beyond the simplifying assumptions of dynamical equilibrium and axisymmetry. All we have done in specifying the GP priors is commit to particular mathematical properties of these quantities (infinitely differentiable and correlations over one or more unknown length-scales). The specific functional form of the velocity moments are updated in the Bayesian sense by information present in the data. \newline

Despite these strengths, the current rendition of our method has an obvious drawback. The lack of a rigorous approach for extracting the uncertainty in predictions for the diagonals of the dispersion tensor makes it difficult to fully interpret any subsequent analysis. There is strong ambiguity in deciding both when to terminate the procedure and how to fragment the dataset. A dedicated investigation into this aspect of the inference pipeline is something that we delegate for future work. However, this shortcoming does not detract from the utility of our method as a general regression tool in the big data regime. \newline

We now return to our earlier assumption made in Section \ref{sec:gaiadata}. We recognize that although recent work indicates that disequilibrium kinematic signatures of solar neighborhood stars are largely independent of color \citep{Bennett2018}, it may be the case that stellar kinematics exhibit non-trivial correlations with other Gaia-derived parameters influenced by the survey's selection function.  For example, studies of the $z - v_z$ phase spiral have revealed its presence in other parameter spaces such as age and metallicity \citep{Frankel2024}. This suggests that kinematic quantities from Gaia DR3 may perhaps be sensitive to sample incompleteness and the specific cuts applied to the color–magnitude diagram. However, since a detailed quantification of the kinematic behavior in the solar neighborhood is not the primary focus of this work, we continue to adopt our initial assumption. Our goal is to demonstrate the applicability of our method as a data regressor, and this assumption does not detract from our conclusions. For future investigation, one might consider extending this analysis to a higher-dimensional space (incorporating additional parameters like age and metallicity), allowing the latent velocity field to vary with both position and these extra degrees of freedom, which could uncover non-trivial dependencies between kinematics and other stellar properties. \newline

Finally, it is worth mentioning that a thorough study of moment equations necessarily involves smooth three dimensional renditions of the velocity field and velocity dispersion tensor, including off diagonal elements. This suggests the natural extension of the present work where we develop the method further to produce a single smooth posterior GP for the first and second velocity moments of the distribution function. This rendition of the method would perform inference on 3-dimensional velocity measurements, allowing us to capture correlations in the kinematics in a bulk region of the Galaxy. A tricky quantity to infer from astrometric data due to selection effects is the zeroth moment of the DF, which describes the number density of stars in the Galaxy \citep{Luri2018}. Moving to a fully 3D rendition of the method would benefit strongly from the ability to infer the smooth density distribution of a stellar sample alongside the velocity moments. Sample completeness is something we did not touch on here, but will become especially important to consider in extending the method to the fully 3D regime. \newline

While our method has been formulated within the context of inferring smooth moments of the stellar DF, we would like to highlight our method's general applicability in an interdisciplinary context. Our SVGPR inference pipeline can be applied to any one-dimensional dataset in astronomy or beyond whose statistical structure adheres to that of equation \ref{eq:1dobswithinputdependentnoisee}. To highlight our method's utility as a general tool for performing non-parametric regression and noise inference, a tutorial for applying the method in a general context will be developed and published online. 

\section{Conclusion} \label{sec:conclusion}

In this paper, we present an GPR algorithm for building data driven models for the first and second velocity moments of the stellar DF. In particular, our method allows one to construct these quantities as a function of a single spatial coordinate in the Galaxy. These models are non-parametric, being built from a statistical distribution placed on the space of possible functions. The mathematical properties of this function space are encoded in a set of mean and covariance functions. \newline

Standard GPR has a few drawbacks that make regression on astrometric data difficult. In particular, it does not possess the computational scalability for the very large data sets encountered in this field and does not provide a way to infer the input dependent noise profile of a dataset alongside the latent process. We have demonstrated that the framework of SVGPR solves the issue of computational complexity and allows us to additionally model and infer the velocity dispersion and mean velocity simultaneously. \newline

Our method provides an attractive alternative to constructing velocity moments from binning procedures, especially if one wishes to make a direct connection between astrometric data and the theoretical framework of Galactic dynamics. The property of our models that bridge the theoretical and observational regimes is the smoothness and differentiability of the resulting velocity moments and paves the way towards a full application of the Jeans framework in studying the dynamics of the Milky Way. Future advancements of our method combined with additional data from future surveys will open the door to exciting analysis possibilities. A 3D rendition of the method that also allows for inference of the smooth stellar density field would allow us to solve the continuity equation in a particular region of the Galaxy, providing direct information about the time dependence of $\rho$. In the data driven era of Galactic dynamics, SVGPR serves as a promising direction for discerning the state of the Milky Way.

\begin{acknowledgments}
This work has made use of data from the European Space Agency (ESA) mission
{\it Gaia} (\url{https://www.cosmos.esa.int/gaia}), processed by the {\it Gaia}
Data Processing and Analysis Consortium (DPAC,
\url{https://www.cosmos.esa.int/web/gaia/dpac/consortium}). Funding for the DPAC
has been provided by national institutions, in particular the institutions
participating in the {\it Gaia} Multilateral Agreement. T.H. and L.M.W. were supported by a Discovery Grant from the Natural Sciences and Engineering Research Council of Canada. T.H. received additional support from the Ontario Graduate Scholarship Program. The Flatiron Institute is funded by the Simon's Foundation.
\end{acknowledgments}

\section*{Data Availability}
The data used in this article to produce 3D stellar velocities is available in the Gaia Data Release 3 at \url{https://gea.esac.esa.int/archive/}. The subset used in this analysis corresponds to stars that survive the cuts described in Section~\ref{sec:gaiadata}.

\facility{Gaia}
\software{GALPY \citep{Bovy2015}, GPyTorch \citep{gardner2021}, tinygp \citep{Foreman-Mackey2021}}

\appendix
\section{Mock Data Tests} \label{app:gridsearchresults}

Here we present the results of our grid search exploring the algorithmic parameter space of our SVGPR setup. Details on the construction of our mock data set used for testing are in section \ref{sec:mockdatatests}. The relevant parameters that we explore in our grid search are the number of training epochs, the batch ratio of SGD, and the independent learning rates of both the latent and dispersion processes. The parameter combinations that we explore are provided in table \ref{tab:paramgrid} below. The GPyTorch framework requires us to select an appropriate treatment for the variational distributions of the approximate GP models representing $\overline{v}_z$ and $\sigma_{\overline{v}_z}$. The two available choices are "CholeskyVariationalDistribution" (CVD) and "NaturalVariationalDistribution" (NVD), and each dictates how hyperparameter updates for the mean and covariance matrices are treated by the optimizer. The former is theoretically motivated for standard optimizers (including SGD), and the latter is designed for use with an optimizer that implements natural gradient descent (NGD). In practice, each optimization problem is uniquely determined by the data set being analyzed, and what is theoretically more appropriate does not apply ubiquitously. In the early stages of this work, we trialed each option both with standard optimizer algorithms (including ADAM) and with GPyTorch's natural gradient optimizer. We found that adopting CVD yielded poor convergence for our particular problem regardless of the optimizer employed. Similarly, adopting NVD paired with NGD also yielded poor results. We saw best performance using NVD in tandem with SGD, which led us to adopt this optimization strategy for all subsequent testing and analysis. In future work, it will be important to focus on a systematic investigation of the interaction between our choices of optimization strategy and parameter grid. \newline

\begin{table}[htbp]
    \centering
    \caption{Values of the SGD optimization parameters that we use in our grid search.}
    \label{tab:paramgrid}
    \begin{tabular}{|c|c|c|c|}
        \hline
        \textbf{Batch Ratio} & \textbf{Latent LR} & \textbf{Dispersion LR} & \textbf{Epochs} \\
        \hline
        20.0 & 0.01 & 0.01 & 300 \\
        50.0 & 0.1 & 0.1 & 400 \\
        100.0 & 1.0 & 1.0 &  \\
        \hline
    \end{tabular}
\end{table}

\begin{figure}[H]
\centering
\includegraphics[width=0.8\textwidth]{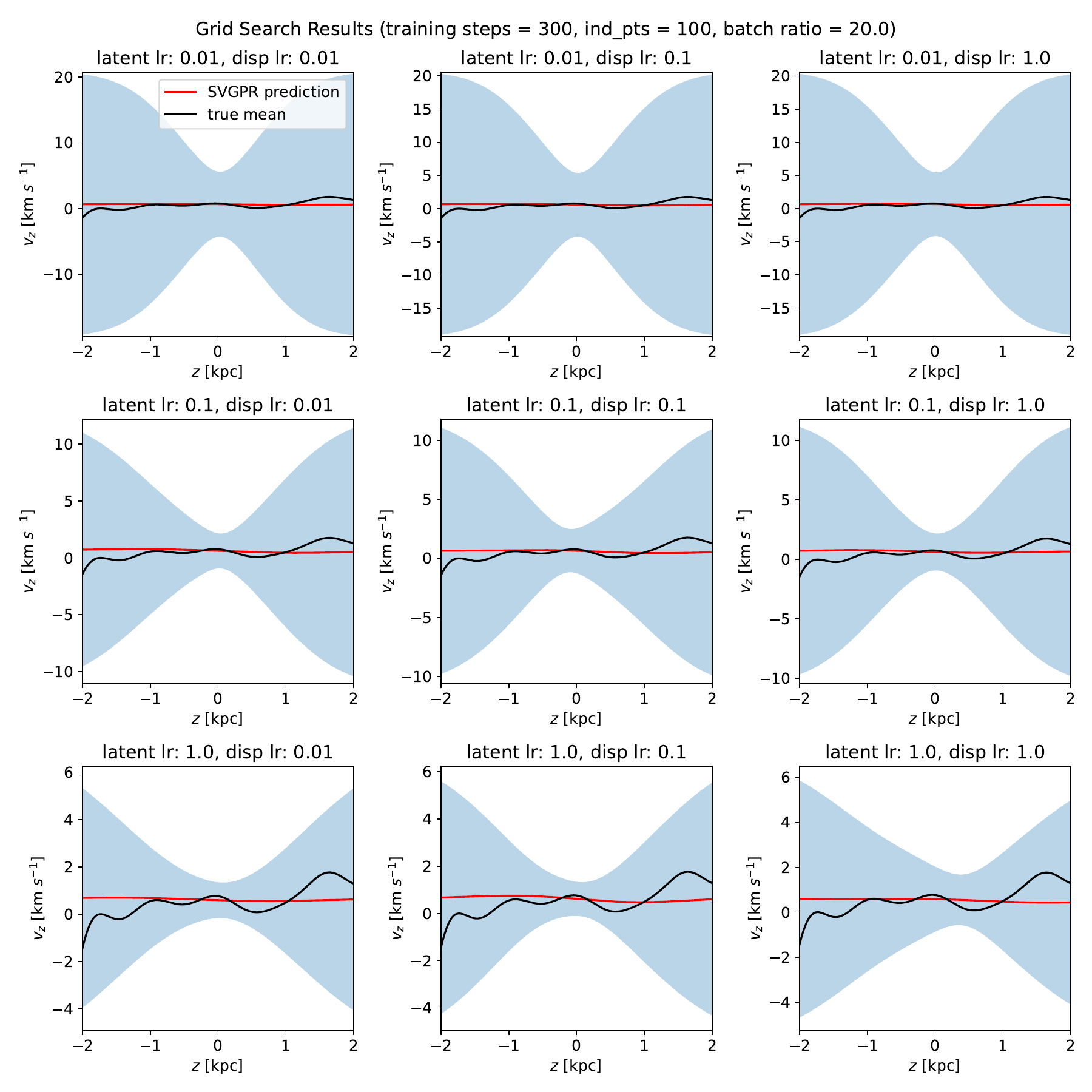}
\caption{SVGPR fits to our mock latent vertical velocity field presented in the left column of figure \ref{fig:mockvzdata} for 300 training steps, a batch ratio of 20.0, and all combinations of learning rate values presented in table \ref{tab:paramgrid}. The red curves depict the SVGPR prediction for the latent velocity, the black curves depict the ground truth, and the blue bands represent the 95 \% confidence regions for the predictions.}
\label{fig:mocklatentbr20}
\end{figure}

\begin{figure}[H]
\centering
\includegraphics[width=0.8\textwidth]{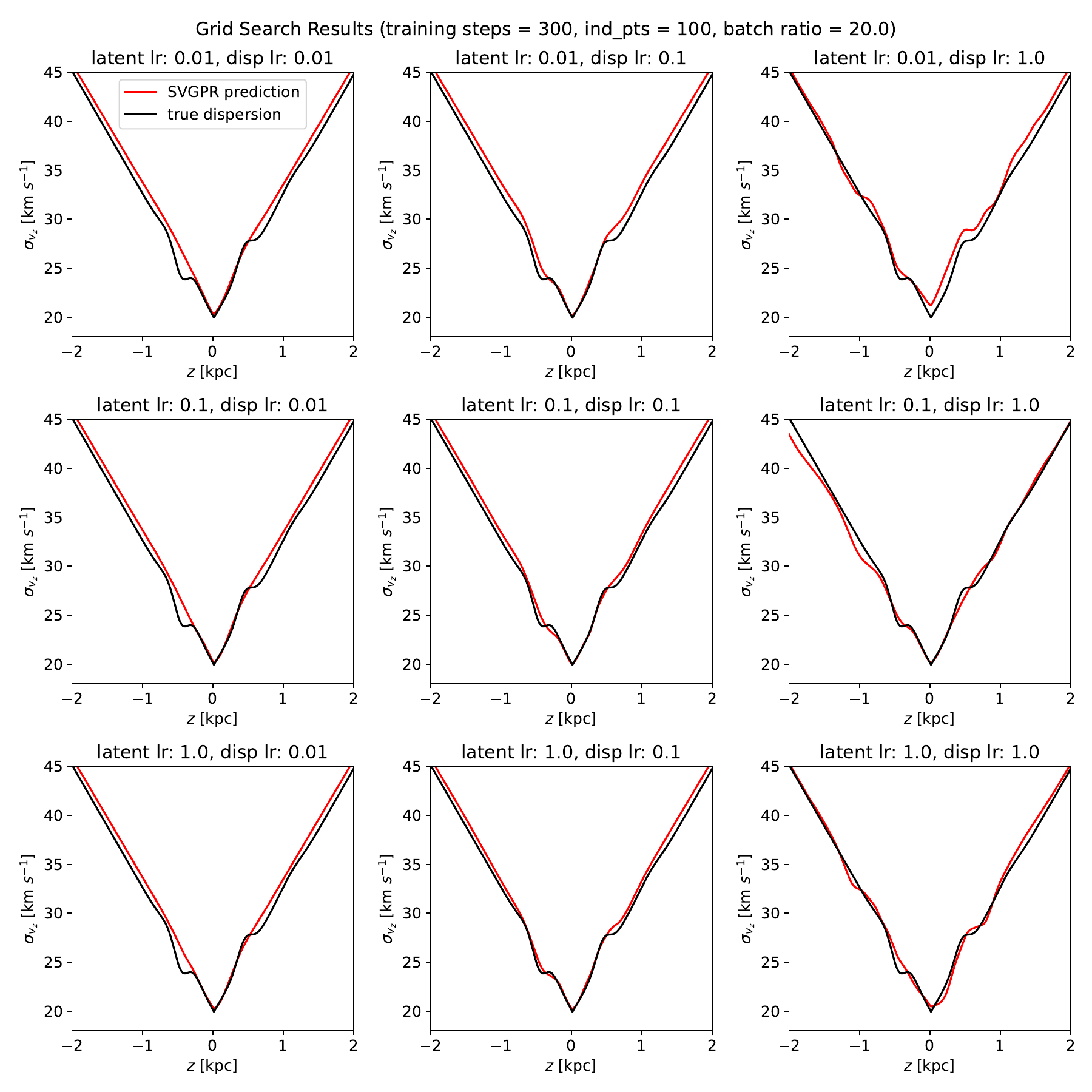}
\caption{SVGPR fits to our mock vertical velocity dispersion profile presented in the right column of figure \ref{fig:mockvzdata} for 300 training steps, a batch ratio of 20.0, and all combinations of learning rate values presented in table \ref{tab:paramgrid}. The red curves depict the SVGPR prediction for the dispersion and the black curves depict the ground truth.}
\label{fig:mockdispbr20}
\end{figure}

\begin{figure}[H]
\centering
\includegraphics[width=0.8\textwidth]{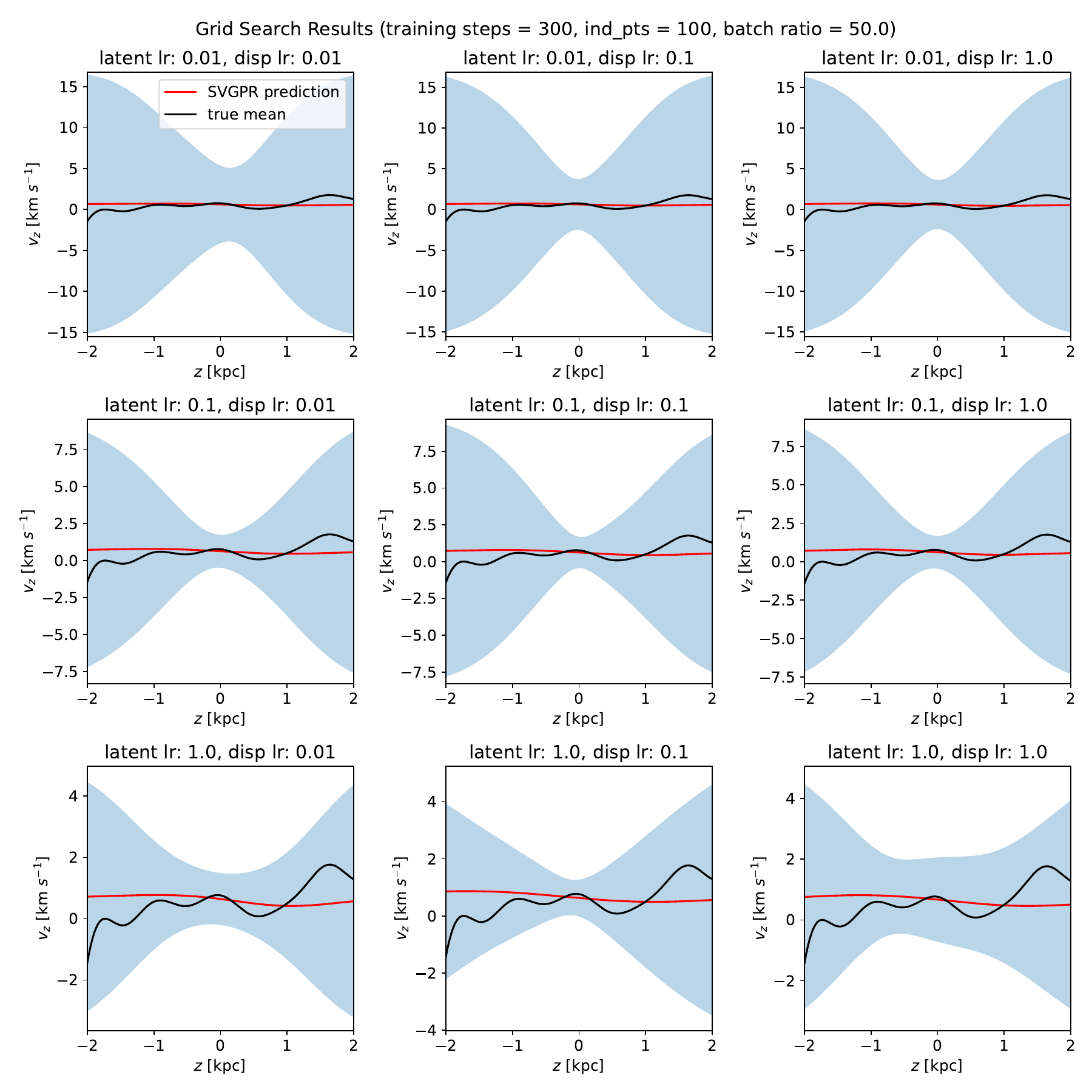}
\caption{Same as in figure \ref{fig:mocklatentbr20}, but this time for a batch ratio of 50.0}
\label{fig:mocklatentbr50}
\end{figure}

\begin{figure}[H]
\centering
\includegraphics[width=0.8\textwidth]{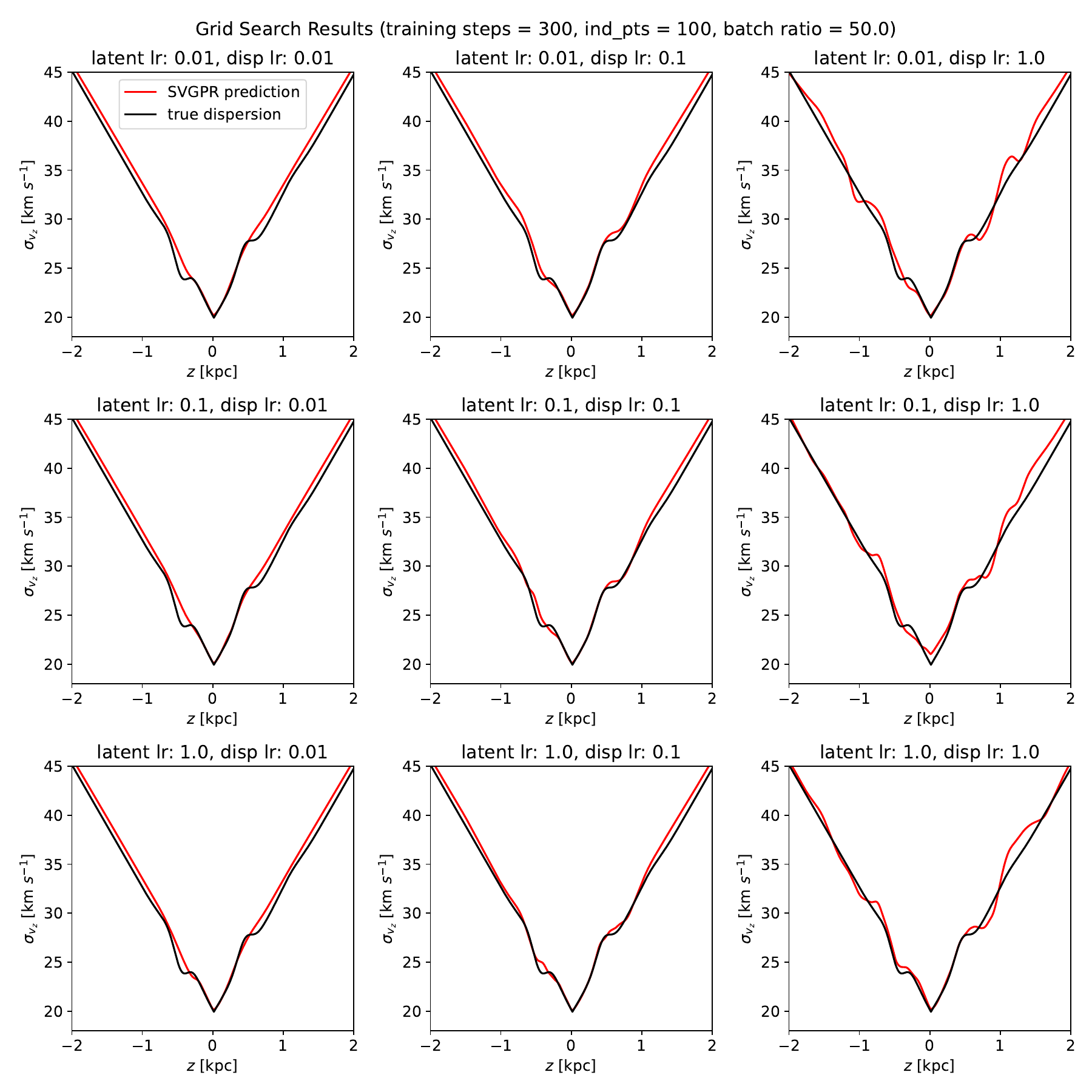}
\caption{Same as in figure \ref{fig:mockdispbr20}, but this time for a batch ratio of 50.0}
\label{fig:mockdispbr50}
\end{figure}

\begin{figure}[H]
\centering
\includegraphics[width=0.8\textwidth]{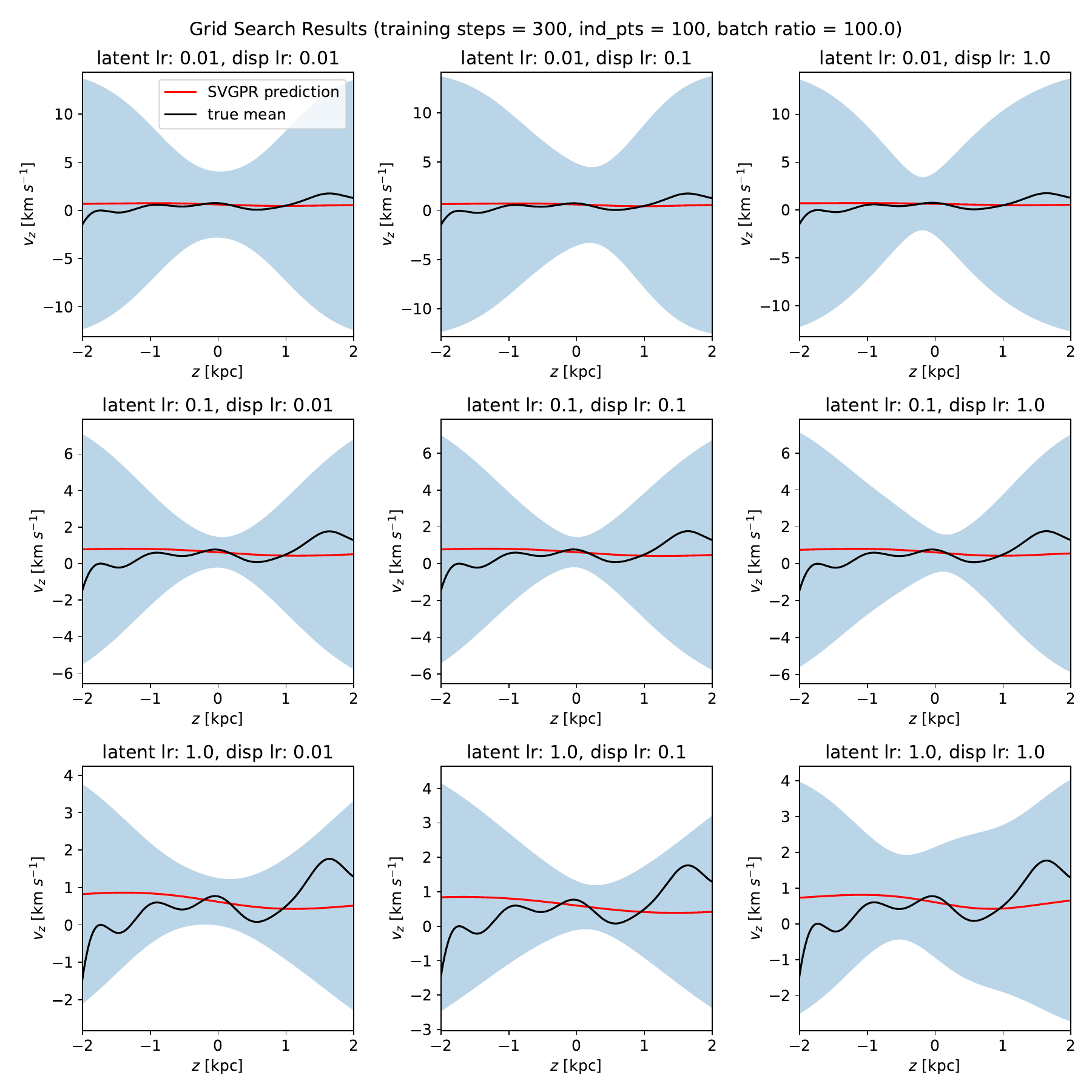}
\caption{Same as in figure \ref{fig:mocklatentbr20}, but this time for a batch ratio of 100.0}
\label{fig:mocklatentbr100}
\end{figure}

\begin{figure}[H]
\centering
\includegraphics[width=0.8\textwidth]{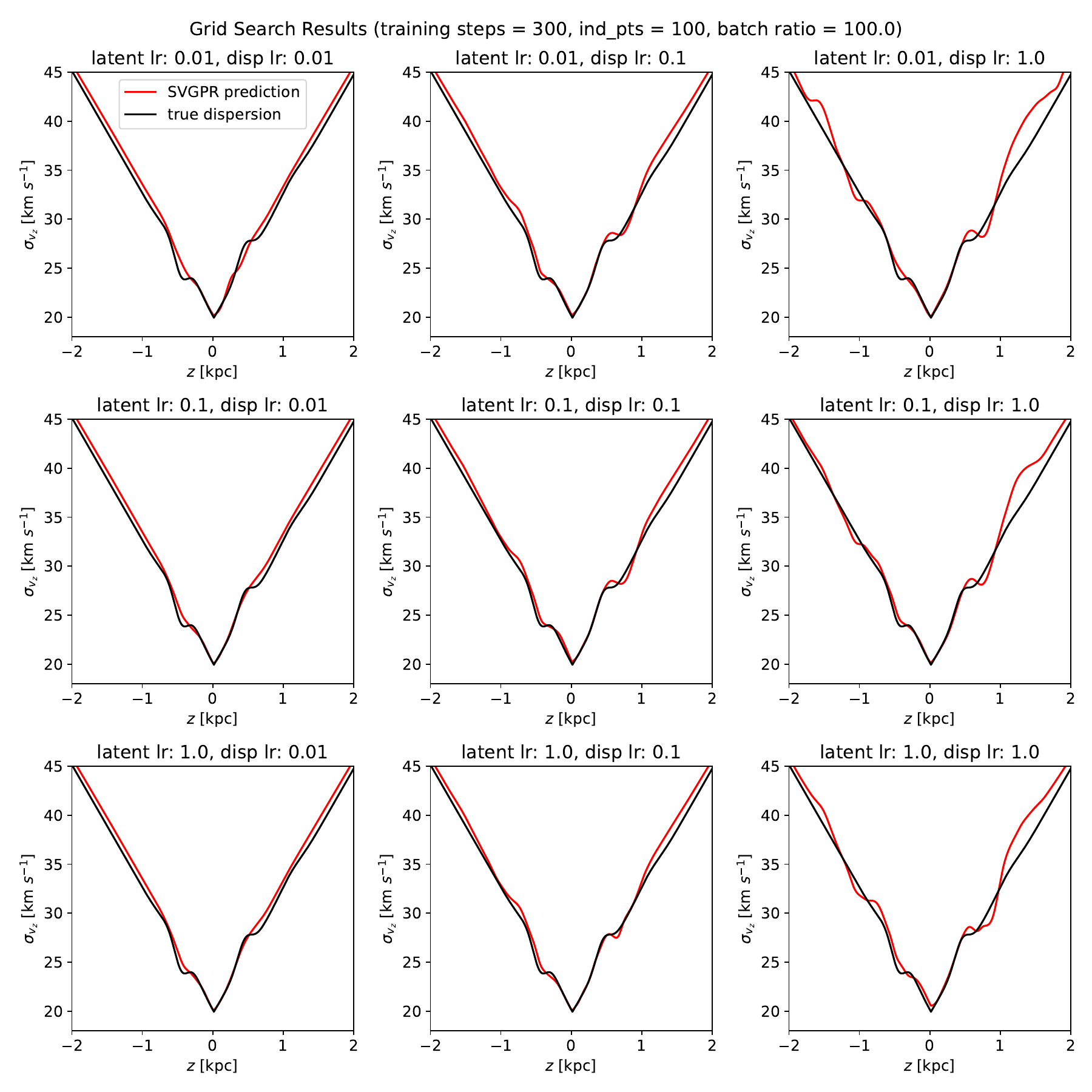}
\caption{Same as in figure \ref{fig:mockdispbr20}, but this time for a batch ratio of 100.0}
\label{fig:mockdispbr100}
\end{figure}

In our grid search, we chose to hold the number of inducing points fixed at $M = 100$. This choice was made both to facilitate computational tractability of the search, and because our primary focus is determining the optimal settings for SGD. Since the inducing points play a different role in the inference procedure than the optimizer parameters, a change in the outcome of regression with $M$ represents an independent degree of freedom. We explore the effects of varying the number of inducing points in the main text, where we adopt the best set of learning rates and training epochs from the results presented here. We carry out our grid search by executing our inference pipeline discussed in \ref{sec:inferenceprocedure} for each combination of values in table \ref{tab:paramgrid}. The resulting fits are provided in figures \ref{fig:mocklatentbr20} through \ref{fig:mockdispbr100}.

\begin{table}[htbp]
    \centering
    \caption{MSE fit results for each learning rate combination (for a batch ratio of 50 and 300 training steps). }
    \label{tab:lr_results_br20}
    \begin{tabular}{|c|c|c|c|}
        \hline
        \textbf{Latent LR} & \textbf{Dispersion LR} & \textbf{Latent MSE} & \textbf{Dispersion MSE} \\
        \hline
        0.01 & 0.01 & 4.591 & 0.958 \\
        0.01 & 0.10 & 4.626 & 0.781 \\
        0.01 & 1.00 & 4.620 & 0.760 \\
        \hline
        0.10 & 0.01 & 4.790 & 0.795 \\
        0.10 & 0.10 & 4.637 & 0.523 \\
        0.10 & 1.00 & 4.715 & 0.540 \\
        \hline
        1.00 & 0.01 & 4.675 & 0.772 \\
        1.00 & 0.10 & 4.660 & 0.449 \\
        1.00 & 1.00 & 4.660 & 0.303 \\
        \hline
    \end{tabular}
\end{table}

\begin{table}[htbp]
    \centering
    \caption{MSE fit results for each learning rate combination (for a batch ratio of 50 and 300 training steps).}
    \label{tab:tab:lr_results_br50}
    \begin{tabular}{|c|c|c|c|}
        \hline
        \textbf{Latent LR} & \textbf{Dispersion LR} & \textbf{Latent MSE} & \textbf{Dispersion MSE} \\
        \hline
        0.01 & 0.01 & 4.650 & 0.718 \\
        0.01 & 0.10 & 4.645 & 0.620 \\
        0.01 & 1.00 & 4.682 & 0.701 \\
        \hline
        0.10 & 0.01 & 4.749 & 0.637 \\
        0.10 & 0.10 & 4.771 & 0.496 \\
        0.10 & 1.00 & 4.746 & 0.806 \\
        \hline
        1.00 & 0.01 & 4.774 & 0.606 \\
        1.00 & 0.10 & 4.953 & 0.426 \\
        1.00 & 1.00 & 4.813 & 0.542 \\
        \hline
    \end{tabular}
\end{table}

\begin{table}[htbp]
    \centering
    \caption{MSE fit results for each learning rate combination (for a batch ratio of 100 and 300 training steps).}
    \label{tab:lr_results_br100}
    \begin{tabular}{|c|c|c|c|}
        \hline
        \textbf{Latent LR} & \textbf{Dispersion LR} & \textbf{Latent MSE} & \textbf{Dispersion MSE} \\
        \hline
        0.01 & 0.01 & 4.700 & 0.605 \\
        0.01 & 0.10 & 4.657 & 0.659 \\
        0.01 & 1.00 & 4.748 & 1.900 \\
        \hline
        0.10 & 0.01 & 4.852 & 0.534 \\
        0.10 & 0.10 & 4.864 & 0.545 \\
        0.10 & 1.00 & 4.782 & 1.198 \\
        \hline
        1.00 & 0.01 & 4.897 & 0.537 \\
        1.00 & 0.10 & 5.068 & 0.501 \\
        1.00 & 1.00 & 4.780 & 1.351 \\
        \hline
    \end{tabular}
\end{table}

Figures \ref{fig:mocklatentbr20}, \ref{fig:mocklatentbr50}, and \ref{fig:mocklatentbr100} display the results for the mock latent velocity fits, and figures \ref{fig:mockdispbr20}, \ref{fig:mockdispbr50}, and \ref{fig:mockdispbr100} show the fits for the dispersion process. In each of these figures, we show all the combinations of learning rates explored for a fixed combination of the SGD batch ratio and the number of training steps. Rows from top to bottom depict increasing latent process learning rate, and columns from left to right correspond to increasing dispersion learning rate. We supplement these figures with the mean squared error values for each fit, which we provide in tables \ref{tab:lr_results_br20} through \ref{tab:lr_results_br100}. \newline

Examining trends across different values of the batch ratio parameter reveals several key insights about the behaviour of optimization. Focusing on the fits in figure \ref{fig:mocklatentbr20}, we notice that as we increase the latent learning rate, the model does not seem to improve in its ability to discern the ground truth profile. What does improve is the size of the confidence region. Moving across the columns from left to right, we see negligible variation in the fit quality, indicating minimal dependence on the dispersion learning rate. For the dispersion process (figure \ref{fig:mockdispbr20}), we do see slight variation when the latent learning rate is increased. The most significant change comes from increasing the dispersion learning rate. The best fit in this figure appears to be the one shown in the middle panel of the bottom row, corresponding to a latent learning rate of 1.0 and a dispersion learning rate of 0.1. Moving one panel over to the right, we see the fit deteriorate again, indicating that for dispersion learning rates greater than 0.1, the algorithm is learning the dispersion process too quickly. This implies that $\sigma_{\overline{v}_z}$ benefits from more conservative learning rates compared to the latent process. \newline

The trends we have observed in figures \ref{fig:mocklatentbr20} and \ref{fig:mockdispbr20} are echoed across all explored values of the batch ratio. Moving onto figures \ref{fig:mocklatentbr50} and \ref{fig:mockdispbr50}, we see the same reduction in the size of the confidence region (which is more drastic than the case of a batch ratio of 20). We also see the same dependence on the dispersion fits with dispersion learning rate, with the algorithm favouring once again the learning rate combination in the middle panel of the bottom row. A glance at the corresponding MLE values for this panel suggests that for increased batch ratio the algorithm performs worse, but this is because the MLE values depend on the discrepancies between the GP mean function (characterized by equation \ref{eq:tanhfit}) and the ground truth for all values of z, not just for those where the features we wish to discern are present. For the latent process, there is now a clear dependence on the latent fit quality with increased latent learning rate. The fact that this behavior appears for a larger batch ratio indicates that the reduction of the number of data points per mini-batch is combating some of the noise of the ELBO landscape, allowing the optimizer to explore alternative minima. Finally, figures \ref{fig:mocklatentbr100} and \ref{fig:mockdispbr100} echo the behaviour seen the previous figures.  \newline

Our grid search also explored training both processes for 400 steps, and we choose not to present all of the results here as we saw little variation compared to the fits already shown. We do however provide the latent and dispersion fits for the best learning rate combination (figure \ref{fig:mockfits400steps} below) to illustrate this, which we deem to be those corresponding to the middle panel in the bottom rows of figures \ref{fig:mockdispbr20}, \ref{fig:mockdispbr50}, and \ref{fig:mockdispbr100}. In fact, for 400 training steps, the dispersion fit is slightly worse than the same fits for 300 training steps, indicating that we are potentially over-training the model. Based on the above results, we conclude that the best algorithmic parameters to adopt are 300 training steps, a batch ratio of 100, a latent learning rate of 1.0 and a dispersion learning rate of 0.1. \newline

\begin{figure}[H]
\centering
\includegraphics[width=0.8\textwidth]{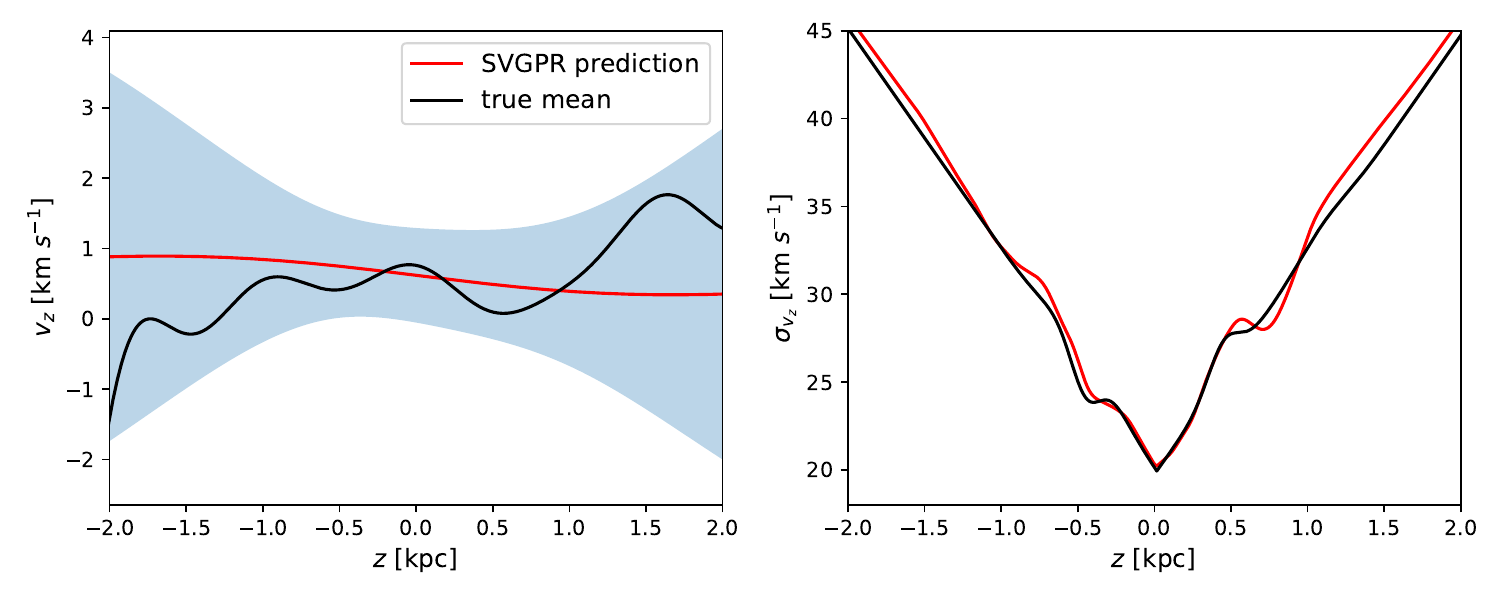}
\caption{SVGPR fit results for our mock latent velocity and velocity dispersion profiles obtained used a batch ratio of 100.0, 400 training steps, a disp}
\label{fig:mockfits400steps}
\end{figure}

These results indicate a disparity between the optimization behaviour of the latent and dispersion processes. The fact that the latent fits show minimal dependence with dispersion learning rate and that the dispersion fits only exhibit weak dependence on the latent learning rate may be indicative of independent treatment of both processes by the optimizer. Furthermore, the fact that the latent process fits change negligibly in moving from 300 to 400 training steps but the dispersion fits slightly worsen may indicate that both GPs could benefit from being trained for a different number of epochs. In particular, these results could mean the latent process benefits from a larger training duration than the dispersion. To explore these possibilities, we perform an additional test. Adopting the best parameter values listed above, we first perform SGD on only the hyperparameters of the latent GP model, holding the dispersion hyperparameters fixed at their initialization values. This is done for double the number of training steps (600). Once latent training is complete, we disconnect the latent GP hyperparameters from the optimizer and train only the dispersion hyperparameters for 300 training steps. Unfortunately, we found no improvement in the model's ability to discern features in the latent process using this approach. The failure to fit both processes to the same degree of precision is discussed in Section \ref{sec:mockdatatests}. \newline

Finally, we note that the fit quality for the same set of parameters appears better in section \ref{sec:mockdatatests} than in the results presented here. In all of our grid search tests, we have artificially reduced the total number of observations in our mock dataset by a factor of 8, while preserving the relative density of data points as a function of $z$. This reduction scheme significantly reduces the computation time of our grid search and retains the same order of magnitude size of a mini-batch of data, but ultimately provides the optimizer with less total information per mini-batch to determine the optimal set of model hyperparameters. This difference in data set size explains the improved performance shown in the main text, and is in line with the expected behaviour that GPR methods perform better with a larger set of measurements \citep{Rasmussen2005-dl}. \newline

\vspace{5mm}

\bibliography{SVGPMilkyWay}
\bibliographystyle{aasjournal}

\end{document}